\documentclass[12pt]{article}
\usepackage{amsfonts,amsmath,amsbsy,amssymb,amsthm,bm,graphicx,latexsym,graphics,epsfig,subfigure,color}
\usepackage{mathrsfs}
\usepackage{longtable,color}
\usepackage{multirow,epsf}
\usepackage{setspace}
\usepackage{float,wasysym}
\usepackage{color}
\usepackage{enumerate} 
\usepackage[dvipsnames]{xcolor}
\usepackage{natbib} \bibpunct{(}{)}{;}{a}{,}{,} 
\usepackage[normalem]{ulem}
\usepackage[colorlinks=true,urlcolor=blue,citecolor=blue,linkcolor=blue,bookmarks=true]{hyperref}
\usepackage{tikz}
\usepackage{multicol}
\usepackage[T1]{fontenc}
\usepackage[utf8]{inputenc}
\def\qed{\hfill \vrule height 6pt width 6pt depth 0pt\\}

\makeatletter
\newcommand*\rel@kern[1]{\kern#1\dimexpr\macc@kerna}
\newcommand*\widebar[1]{%
  \begingroup
  \def\mathaccent##1##2{%
    \rel@kern{0.8}%
    \overline{\rel@kern{-0.8}\macc@nucleus\rel@kern{0.2}}%
    \rel@kern{-0.2}%
  }%
  \macc@depth\@ne
  \let\math@bgroup\@empty \let\math@egroup\macc@set@skewchar
  \mathsurround\z@ \frozen@everymath{\mathgroup\macc@group\relax}%
  \macc@set@skewchar\relax
  \let\mathaccentV\macc@nested@a
  \macc@nested@a\relax111{#1}%
  \endgroup
}
\makeatother

\allowdisplaybreaks

\newtheorem{The}{Theorem}

\newtheorem{Lem}[The]{Lemma}
\newtheorem{Cor}[The]{Corollary}
\newtheorem{Def}[The]{Definition}
\newtheorem{Exa}[The]{Example}
\newtheorem{Rem}[The]{Remark}

\numberwithin{The}{section}

\newcommand{\dif}{\mathrm{d}}

\begin{document}

\title{\vskip -3cm
\bf Systemic Risk:\\ Conditional Distortion Risk Measures}
\author{Jan Dhaene\\
{\footnotesize Faculty of Business and Economics}\\
{\footnotesize Katholieke Universiteit Leuven}\\
{\footnotesize {\tt Jan.Dhaene@kuleuven.be}}\\\and Roger J. A. Laeven\\
{\footnotesize Amsterdam School of Economics}\\
{\footnotesize University of Amsterdam, KU Leuven}\\
{\footnotesize and CentER}\\
{\footnotesize {\tt R.J.A.Laeven@uva.nl}}\\\and Yiying Zhang\\
{\footnotesize School of Statistics and Data Science, LPMC and KLMDASR}\\
{\footnotesize Nankai University, Tianjin 300071, P. R. China}\\
{\footnotesize {\tt zhangyiying@outlook.com}}}
\date{\today}
\maketitle

\begin{abstract}
In this paper, we introduce the rich classes of conditional distortion (CoD) risk measures and distortion risk contribution ($\Delta$CoD) measures as measures of systemic risk
and analyze their properties and representations.
The classes include the well-known conditional Value-at-Risk, conditional Expected Shortfall,
and risk contribution measures in terms of the VaR and ES as special cases.
Sufficient conditions are presented for two random vectors to be ordered by the proposed CoD-risk measures and distortion risk contribution measures. These conditions are expressed using the conventional stochastic dominance, increasing convex/concave, dispersive, 
and excess wealth orders of the marginals 
and canonical positive/negative stochastic dependence notions.
Numerical examples are provided to illustrate our theoretical findings.  
This paper is the second in a triplet of papers on systemic risk by the same authors.  
In \cite{DLZorder2018a}, we introduce and analyze some new stochastic orders related to systemic risk. 
In a third (forthcoming) paper, we attribute systemic risk to the different participants in a given risky environment.\\[3mm]
\end{abstract}

\newpage

\section{Introduction}
Risk measures are commonly used as capital requirements,
i.e., as real-valued mappings on a class of financial positions to determine the amount of risk capital
to be held in reserve. 
The purpose of this risk capital is to make the risk given by the financial position
taken by a financial institution, such as an insurance company or a bank,
acceptable from a \textit{microprudential} regulatory perspective.
Prominent examples of risk measures are the Value-at-Risk (VaR) and the Expected Shortfall (ES).\footnote{Expected Shortfall is
also referred to as Tail Value-at-Risk (TVaR) and Average Value-at-Risk (AVaR);
see Definition \ref{defvacva} for the explicit definitions.}
Indeed, with the international adoption of financial regulatory frameworks
such as the Basel Capital Accords for banks and the European Solvency Regulation for insurers,
VaR has become the predominant measure of risk for financial institutions.
Since its first adoption in the nineties of the previous century,
an active area of research has analyzed the appealing and appalling properties of VaR (and, later, ES),
and has developed new alternative theories of risk measurement.
These theories build on a rich literature in actuarial mathematics and decision theory and have nowadays reached
a high level of mathematical and economic sophistication.

Over the past decade we have witnessed several occurrences of pronounced transmissions of
adverse economic events within a highly interconnected financial-economic network,
at an international or even global scale.
The interactions among risks in the form of stochastic interdependences,
in part caused by dynamic feedback relations,
should play a central role in quantitative risk analysis;
see \cite{Denuit2005}, \cite{Embrechts2005}, \cite{Kaas2009}, \cite{Laeven2009}, and \cite{Goovaerts2011} in the context of risk aggregation
under VaR and ES.
Since the 2008/09 global financial crisis, risk measures have increasingly been employed not just
to provide microprudential assessments of marginal risks or aggregate portfolio risks,
but also to evaluate forms of systemic risk:
from a \textit{macroprudential} perspective we are interested in
the systemic risk that a failure or loss of one entity spreads contagiously to other entities
or even to the entire financial system.
Indeed, the complex system of financial institutions in a competitive economy induces an undeniable presence of interconnectedness.
This interconnectedness can cause a collapse in part of the system as a result of a contagious disruption due to the failure of a singular player.
Thus, the potential threat of the failure of a singular player can have a reverberating effect on the security and stability of the system
and the economy as a whole.
In the literature, several papers have proposed different conditional risk (co-risk) measures and risk contribution measures
to evaluate the systemic risks emerging from a group of financial institutions and the interactions among them;
see e.g., \cite{Gourieroux2013}, \cite{Girardi2013}, \cite{Adrian2016}, \cite{brownlees2016}, \cite{Acharya2010}, and the references therein.

As a simple measure of systemic risk, \cite{Adrian2016} analyze the conditional VaR
(CoVaR, see Definition \ref{defCoVaR}).
It is defined as the VaR of one specific financial institution,
conditional upon the occurrence of an event that is specific to another financial institution.
The prefix ``Co'' is meant to refer to ``conditional'' (or ``co-movement'')
and emphasizes the systemic nature of this measure of risk.
In a sense, CoVaR provides a measure of a spillover effect.
In related literature, \cite{mainik2014} introduce the conditional expected shortfall (CoES)
and \cite{Acharya2010} propose the marginal expected shortfall (MES);
see Definitions \ref{defCoES} and \ref{defMES}.
For a given choice of such co-risk measures, the associated risk contribution measure
evaluates how a stress scenario for one component incrementally affects another component or the entire system.
Examples of risk contribution measures including $\Delta$CoVaR (Definition \ref{defconcovar}) and $\Delta$CoES (Definition \ref{defconcoes})
can be found in \cite{Girardi2013}, \cite{mainik2014}, and \citet{Adrian2016}.
Using data on U.S. financial institutions over the period 2005-2014
\cite{Kleinow2017} compare several commonly used systemic risk metrics, including CoVaR and MES.
They illustrate that the alternative measurement approaches produce very different estimates of systemic risk.
In particular, different systemic risk metrics may lead to contradicting assessments
about the riskiness of different types of financial institutions.
As mentioned in \cite{mainik2014} ``\emph{... the dependence consistency or, say, dependence coherency of systemic risk indicators is a novel problem area that needs further study ... questions for general characterizations or representations of functionals ... are currently open.}''

In the context of comparisons of these co-risk measures and risk contribution measures,
an interesting paper by \cite{sordo2018IME} provides sufficient conditions to stochastically order two random vectors
in terms of their CoVaR, CoES, $\Delta$CoVaR, and $\Delta$CoES,
where the conditions are expressed using conventional stochastic orders for the marginals
under some assumptions of positive dependence.
Furthermore, \cite{fang2018RISKS} investigate how the marginal distributions and the dependence structure
affect the interactions among paired risks under the above co-risk measures and risk contribution measures.
It is well known that the VaR and ES arise as two special cases within the rich class of distortion risk measures
\citep[][]{Yaari1987, Denuit2005, Denuit2006, Goovaerts2010, Follmer2011}.
Distortion risk measures satisfy several appealing (in fact, characterizing) properties
including monotonicity, translation invariance, comonotonic additivity, and positive homogeneity.
Besides, distortion risk measures are consistent with the usual stochastic order (i.e., first-order stochastic dominance)
under any distortion function,
and with the increasing convex order (i.e., stop-loss order)
under any concave distortion function.
Furthermore, concave distortion risk measures occur naturally as building blocks of law-invariant convex risk measures
\citep[see Chapter 4 in][]{Follmer2011}.

The aim of this paper is to introduce general and unified classes of conditional risk measures and risk contribution measures
by means of distortion functions.
This gives rise to \textit{conditional distortion (CoD) risk measures} and \textit{distortion risk contribution ($\Delta$CoD) measures}.
We analyze the properties and present representations of these new systemic risk measures.
We establish sufficient conditions for ordering two bivariate random vectors
by the proposed systemic risk measures in terms of the canonical stochastic orders
(e.g., first-order stochastic dominance, the increasing concave/convex order, the dispersive order, and the excess wealth order) between the marginals, dependence structure, distortion functions, and threshold quantiles.
The interactions between paired risks are also investigated.
Existing results in \cite{mainik2014}, \cite{sordo2018IME}, and \cite{fang2018RISKS} are generalized and extended.

In a somewhat related strand of the literature, \cite{Hoffmann2016} axiomatically introduced
risk-consistent conditional systemic risk measures
defined on multidimensional risks.
This class consists of those conditional systemic risk measures
that can be decomposed into a state-wise conditional aggregation
and a univariate conditional risk measure.
Their studies extend known results for unconditional risk measures on finite state spaces.
Besides, \cite{Biagini2018} specified a general methodological framework for systemic risk measures
via multidimensional acceptance sets and aggregation functions.
Their approach yields systemic risk measures that can be given the interpretation of
the minimal amount of cash that safeguards the aggregated system.

The organization of the present paper is as follows.
In Section \ref{sec:pre}, we recall some useful definitions and concepts.
In Section \ref{sec:distordef}, we introduce conditional distortion (CoD) risk measures and distortion risk contribution ($\Delta$CoD) measures,
and give some useful expressions employed in the sequel.
Section \ref{sec:diststo} studies the comparisons of two random vectors under CoD-risk measures,
and provides sufficient conditions for their ordering in terms of the usual stochastic order, the increasing convex order, and the increasing concave order of marginals, under appropriate assumptions on the dependence structure, distortion functions, and threshold quantiles.
In Section \ref{sec:contridis}, we present sufficient conditions for comparison of the distortion risk contribution measures in terms of the dispersive order and the excess wealth order of marginals.
Section \ref{sec:paired} investigates the interactions between paired risks under our proposed new CoD-risk measures and distortion risk contribution measures.
Section \ref{sec:nume} provides some numerical examples to illustrate our main findings.
Section \ref{sec:conclusions} concludes the paper.

\section{Preliminaries}\label{sec:pre}
Throughout this paper, the term ``increasing'' is used for ``non-decreasing'' and ``decreasing''
is used for ``non-increasing''.
Expectations and density functions
are assumed to exist when they appear.
Let $\mathcal{C}$ and $\mathcal{C}^2$ be the set (space) of all univariate and bivariate distribution functions considered in the sequel.
Furthermore, let $\mathbb{R}$ be the set of real numbers and $\mathbb{R}_{\geq 0}=[0,+\infty)$,
and let $\mathbb{N}_+$ be the set of strictly positive natural numbers.
We use the expression `$X\sim F$' to denote that the random variable (or vector) has distribution $F$,
and use $\underline{X}$ to denote the random vector $(X_1,\ldots,X_n)$.

\subsection{Stochastic Orders}
We denote by $X$ and $Y$ two random variables (r.v.'s) with respective distribution functions (d.f.'s) $F$ and $G$,
survival functions $\widebar{F}$ and $\widebar{G}$, and density functions $f$ and $g$.
Let $F^{-1}(p)=\inf\{x\in\mathbb{R}\mid F(x)\geq p\}$ and $G^{-1}(p)=\inf\{x\in\mathbb{R}\mid G(x)\geq p\}$ be the generalized inverses of 
the d.f.'s $F$ and $G$ of $X$ and $Y$, for $p\in[0,1]$, respectively, where $\inf\emptyset=+\infty$ by convention.
\begin{Def} $X$ is said to be smaller than $Y$ in the
\begin{itemize}
\item [(i)] {likelihood ratio order} (denoted by $X\leq_{\rm lr} Y$) if $g(x)/f(x)$ is increasing in $x\in\mathbb{R}$;
\item [(ii)] {hazard rate order} (denoted by $X\leq_{\rm hr} Y$) if $\widebar{G}(x)/\widebar{F}(x)$ is increasing in $x\in\mathbb{R}$;
\item [(iii)] {usual stochastic order} (denoted by $X\leq_{\rm st} Y$) if $\widebar{F}(x)\leq \widebar{G}(x)$ for all $x\in\mathbb{R}$;
\item [(iv)] {increasing convex order} (denoted by $X\leq_{\rm icx} Y$) if $\mathbb{E}[\phi(X)]\leq\mathbb{E}[\phi(Y)]$ for any increasing and convex function $\phi:\mathbb{R}\to\mathbb{R}$;
\item [(v)] {increasing concave order} (denoted by $X\leq_{\rm icv} Y$) if $\mathbb{E}[\phi(X)]\leq\mathbb{E}[\phi(Y)]$ for any increasing and concave function $\phi:\mathbb{R}\to\mathbb{R}$;
\item [(vi)] {dispersive order} (denoted by $X\leq_{\rm disp}Y$) if $F^{-1}(v)-F^{-1}(u)\leq G^{-1}(v)-G^{-1}(u)$, for all $0< u\leq v<1$;
\item [(vii)] {excess wealth order} (denoted by $X\leq_{\rm ew}Y$) if $\int_{F^{-1}(u)}^{\infty}\widebar{F}(t)\dif t\leq \int_{G^{-1}(u)}^{\infty}\widebar{G}(t)\dif t$, for all $u\in(0,1)$.
\end{itemize}
\end{Def}

As is well known,
\begin{equation*}
X\leq_{\rm lr} Y\Longrightarrow X\leq_{\rm hr} Y \Longrightarrow X\leq_{\rm st} Y \Longrightarrow X\leq_{\rm icx~[icv]} Y.
\end{equation*}
Furthermore, the dispersive order is stronger than (i.e., implies) the excess wealth order,
and is a partial order used to compare the variabilities among two probability distributions.
For comprehensive discussions on these useful partial orders,
we refer the reader to the monographs by \cite{Denuit2005}, \cite{marshall2007life}, and \cite{shaked2007sto}.

\subsection{Measuring Dependence}
The following notions entail that, for a bivariate random vector,
larger values of one component are associated with larger values of the other,
in some specific sense.
\begin{Def} \begin{itemize}
\item [(i)] The bivariate random vector $(X,Y)$ is said to be totally positive of order 2 [reverse regular of order 2] (written as {\rm TP}$_{2}$ [{\rm RR}$_2$]) if $[X|Y=y_1]\leq_{\rm lr} [\geq_{\rm lr}]\ [X|Y=y_2]$, for all $y_1\leq y_2$, and $[Y|X=x_1]\leq_{\rm lr} [\geq_{\rm lr}]\ [Y|X=x_2]$, for all $x_1\leq x_2$.
\item [(ii)] $X$ is said to be stochastically increasing [decreasing] in $Y$ (written as $X\uparrow_{\rm SI~[SD]}Y$) if $[X|Y=y_1]\leq_{\rm st}[\geq_{\rm st}][X|Y=y_2]$, for all $y_1\leq y_2$.
\item [(iii)] The bivariate random vector $(X,Y)$ is said to be positively [negatively] dependent through stochastic ordering (PDS~[NDS])
if $X\uparrow_{\rm SI~[SD]}Y$ and $Y\uparrow_{\rm SI~[SD]}X$.
\item [(iv)] $X$ is said to be right tail increasing [decreasing] in $Y$ (written as $X\uparrow_{\rm RTI~[RTD]}Y$) if $\mathbb{P}(X>x|Y>y)$ is increasing in $y\in\mathbb{R}$, for all $x\in\mathbb{R}$.
\item [(v)] The bivariate random vector $(X,Y)$ is said to be positive [negative] quadrant dependent (PQD~[NQD]) if, for all $(x,y)\in\mathbb{R}^2$,
it holds that
\begin{equation*}
\mathbb{P}(X>x,Y>y)\geq[\leq]\mathbb{P}(X>x)\mathbb{P}(Y>y),
\end{equation*}
or equivalently,
\begin{equation*}
\mathbb{P}(X\leq x,Y\leq y)\geq[\leq]\mathbb{P}(X\leq x)\mathbb{P}(Y\leq y).
\end{equation*}
\end{itemize}
\end{Def}

The following implications (with slight abuse of notation) are well known:
\begin{equation*}
\mbox{$(X,Y)$ is TP$_2$}\Longrightarrow X\uparrow_{\rm SI}Y ~[Y\uparrow_{\rm SI}X] \Longrightarrow X\uparrow_{\rm RTI}Y~[Y\uparrow_{\rm RTI}X] \Longrightarrow \mbox{$(X,Y)$ is PQD},
\end{equation*}
\begin{equation*}
\mbox{$(X,Y)$ is RR$_2$}\Longrightarrow X\uparrow_{\rm SD}Y~[Y\uparrow_{\rm SD}X] \Longrightarrow X\uparrow_{\rm RTD}Y~[Y\uparrow_{\rm RTD}X] \Longrightarrow \mbox{$(X,Y)$ is NQD}.
\end{equation*}
It is also clear that if $(X,Y)$ is TP$_2$ [RR$_2$] then it must be PDS [NDS].
Besides, $(X,Y)$ is TP$_2$ [RR$_2$]
if and only if its copula $C$ is TP$_2$ [RR$_2$] \citep[see][]{Muller2002, Cai2012}.
For more detailed discussions, interested readers are referred to \cite{Barlow1975}, \cite{Block1982}, \cite{Joe1997},
and \cite{Denuit2005}.

\subsection{Comparing Dependence and Copulas}
Consider a bivariate random vector $(X,Y)$ with respective marginal d.f.'s $F$ and $G$
and joint d.f. $H$.
It is well known that any such bivariate d.f. $H$ admits the decomposition
\begin{equation*}
H(x,y)=C(F(x),G(y)),\qquad x,y\in\mathbb{R},
\end{equation*}
where $C$ is a bivariate d.f. on $(0,1)^2$ with uniform margins \citep[][]{sklar1959},
i.e., there exist r.v.'s $U,V\sim U(0,1)$ such that $C(u,v)=\mathbb{P}(U\leq u, V\leq v)$.
The function $C$ is called a \emph{copula} of $H$.
If both $F$ and $G$ are continuous, then $C$ is uniquely determined by $C(u,v)=H(F^{-1}(u),G^{-1}(v))$.
The copula characterizes the dependence of the random vector $(X,Y)$.
We also denote by $\widebar{C}$ the \emph{joint tail function}
for two uniform r.v.'s whose joint d.f. is the copula $C$, that is,
\begin{equation*}
\widebar{C}(u,v)=\mathbb{P}(U> u, V> v)=1-u-v+C(u,v),\quad\mbox{$(u,v)\in(0,1)^2$}.
\end{equation*}
The joint tail function $\widebar{C}$ should not be confused with the \emph{survival copula} of $U$ and $V$,
which is defined as
\begin{equation*}
\hat{C}(u,v)=u+v-1+C(1-u,1-v),\quad\mbox{$(u,v)\in(0,1)^2$}.
\end{equation*}
The survival copula $\hat{C}$ couples the joint survival function to its univariate margins (survival functions)
in a manner completely analogous to how a copula links the joint d.f. to its margins.
Clearly, $\widebar{C}(u,v)=\hat{C}(1-u,1-v)$.

We next recall the definition of the concordance order \citep[see Definition 2.8.1 in][]{Nelsen2007}.
\begin{Def} Given two copulas $C$ and $C'$, $C$ is said to be smaller than $C'$ in concordance order
(denoted as $C\prec C'$) if $C(u,v)\leq C'(u,v)$, for all $(u,v)\in(0,1)^2$.
\end{Def}

The concordance order is also referred to as the correlation order or the positive quadrant dependence (PQD) order in the literature \citep[see][]{dhaene1996dependency, dhaene1997dependency, Nelsen2007}.
It is a partial order as not every pair of copulas is concordance-comparable.
Besides, the canonical scale-free dependence measures given by Kendall's tau and Spearman's rho
are well known to be increasing with respect to the concordance order.

\subsection{Distortion Risk Measures}
We state the following definition:
\begin{Def}\label{defvacva}The {\rm VaR} and {\rm ES} of a r.v. $X$ with d.f. $F$
at confidence level $\alpha\in(0,1)$ are defined as
\begin{equation*}
{\rm VaR}_{\alpha}[X]= F^{-1}(\alpha),
\end{equation*}
and
\begin{equation*}
{\rm ES}_{\alpha}[X]= \frac{1}{1-\alpha}\int_{\alpha}^{1}{\rm VaR}_{p}[X]\dif p,
\end{equation*}
provided that the integral exists.
\end{Def}

VaR and ES occur as special cases of distortion risk measures
\citep[][]{Yaari1987, Denuit2005, Denuit2006, Dhaene2006, Goovaerts2010, Follmer2011}.
In full generality, a {\em distortion function} $g: [0,1]\mapsto[0,1]$
is an increasing function such that $g(0)=0$ and $g(1)=1$.
The set of all distortion functions is denoted by $\mathcal{G}$.
A distortion risk measure, then, is a functional mapping the elements of $\mathcal{C}$ to the real line, as follows:
\begin{Def}\label{defdistor}For a distortion function $g\in\mathcal{G}$,
the distortion risk measure $\mbox{{\rm D}}_{g}$ of a r.v. $X$ with d.f. $F$
is defined as
\begin{equation*}
\mbox{{\rm D}}_{g}[X]=-\int_{-\infty}^{0}[1-g(\widebar{F}(t))]\dif t+\int_{0}^{+\infty}g(\widebar{F}(t))\dif t.
\end{equation*}
In particular, if $X$ is a nonnegative r.v., then
\begin{equation*}
\mbox{{\rm D}}_{g}[X]=\int_{0}^{+\infty}g(\widebar{F}(t))\dif t.
\end{equation*}
\end{Def}

It is well known that distortion risk measures enjoy, and can be characterized by,
the properties of monotonicity, translation invariance, comonotonic additivity, and positive homogeneity.
Besides, distortion risk measures are consistent with the usual stochastic order under any distortion function $g\in\mathcal{G}$,
and with the increasing convex order under any concave distortion function.
\cite{Wirch2001} showed that a distortion risk measure is coherent,
i.e., monotonic, translation invariant, positively homogeneous, \textit{and} subadditive
\citep[see e.g.,][]{Follmer2011, Laeven2013}
if and only if the distortion function is concave.
Furthermore, concave distortion functions are the building blocks of law-invariant convex risk measures
\citep[see Chapter 4 in][]{Follmer2011}.

The two prominent examples of distortion risk measures given by ${\rm VaR}_{\alpha}$ and ${\rm ES}_{\alpha}$
correspond to the distortion functions $g(p)={\bm1}_{(1-\alpha,1]}(p)$ and $g(p)=\min\{1,\frac{p}{1-\alpha}\}$, for $\alpha\in(0,1)$, respectively.
Obviously, the distortion function for VaR is not continuous but left-continuous, while the distortion function for ES is continuous and concave but not differentiable everywhere.
Besides, the incomplete beta function, the Wang distortion or Esscher-Girsanov transform,
and the lookback distortion are commonly used special cases of distortion functions; see \cite{Denuit2006} and \cite{Goovaerts2008}.
It is easily verified \citep[see e.g.,][]{Belles2014} that for a r.v. $X$ and any two distortion functions $g,g'\in\mathcal{G}$, $g\leq g'$ implies that $\mbox{{\rm D}}_{g}[X]\leq \mbox{{\rm D}}_{g'}[X]$.

Next, we introduce the notion of a dual distortion function.
Consider a distortion function $g$ and define the related function $\widebar{g}: [0,1]\mapsto[0,1]$ by $\widebar{g}(p)=1-g(1-p)$, for $p\in[0,1]$.
Obviously, $\widebar{g}$ is also a distortion function, called the \emph{dual distortion function} of $g$.
It is well known that, for any r.v. $X$ and distortion function $g$, $\mbox{{\rm D}}_{\widebar{g}}[X]=-\mbox{{\rm D}}_{g}[-X]$ and $\mbox{{\rm D}}_{g}[X]=-\mbox{{\rm D}}_{\widebar{g}}[-X]$
\citep[see Lemma 5 in][]{dhaene2012remarks}.
Note that if $g$ is left-continuous, then $\widebar{g}$ is right-continuous (c.f. Theorems 4 and 6 in \cite{dhaene2012remarks}).

For a right-continuous distortion function $g$, the transformation of the tail function $\widebar{F}=1-F$ of $X$ given by $g(\widebar{F}(x))=g\circ\widebar{F}(x)$ defines a new tail function associated to a r.v. $X_g$, which is the distorted counterpart of the r.v. $X$, 
induced by distorting $X$ with distortion function $g$.

For more discussions on the properties and applications of distortion risk measures, one may refer to \cite{Wang1997}, \cite{hurlimann2004}, \cite{Denuit2005, Denuit2006}, \cite{Dhaene2006}, \cite{balbas2009}, and \cite{dhaene2012remarks}.

\subsection{Co-Risk Measures and Risk Contribution Measures}
Conditional risk (co-risk) measures are increasingly employed as measures of systemic risk.
Prototypical examples of co-risk measures include the conditional Value-at-Risk (CoVaR)
\citep[][]{Adrian2016,Girardi2013},
the conditional Expected Shortfall (CoES)
\citep[][]{mainik2014},
and the marginal Expected Shortfall (MES) \citep[][]{Acharya2010}.
For a given co-risk measure, the corresponding risk contribution measure assesses the incremental effect of a stress scenario.
Well known examples of risk contribution measures include $\Delta$CoVaR and $\Delta$CoES;
see \cite{Girardi2013}, \cite{mainik2014}, and \cite{Adrian2016}.

The definition of CoVaR is given as follows:
\begin{Def}\label{defCoVaR} Let $\alpha, \beta\in(0,1)$. Then,
\begin{equation*}
\mbox{{\rm CoVaR}}_{\alpha, \beta}[Y|X]=\mbox{\rm VaR}_{\beta}[Y|X>\mbox{\rm VaR}_{\alpha}[X]].
\end{equation*}
\end{Def}

The above definition is adapted by \cite{mainik2014} to the case of CoES, under which the coherence of ES is inherited by CoES:
 \begin{Def}\label{defCoES} Let $\alpha, \beta\in(0,1)$. Then,
\begin{equation*}
\mbox{{\rm CoES}}_{\alpha, \beta}[Y|X]=\frac{1}{1-\beta}\int_{\beta}^{1}\mbox{{\rm CoVaR}}_{\alpha, t}[Y|X]\dif t.
\end{equation*}
\end{Def}

One easily verifies that, with continuous marginals,
the CoES can be represented through a conditional expectation of $Y$,
in a manner similar to the familiar representation of ES:
\begin{equation*}
\mbox{{\rm CoES}}_{\alpha, \beta}[Y|X]=\mathbb{E}[Y|X>\mbox{\rm VaR}_{\alpha}[X], Y>\mbox{{\rm CoVaR}}_{\alpha, \beta}[Y|X]].
\end{equation*}

The definition of MES is given as follows:
\begin{Def}\label{defMES} Let $\alpha\in(0,1)$. Then,
\begin{equation*}
\mbox{{\rm MES}}_{\alpha}[Y|X]=\mathbb{E}[Y|X>\mbox{\rm VaR}_{\alpha}[X]].
\end{equation*}
\end{Def}

To measure the risk contribution of $X$ to $Y$, one may compare $\mbox{{\rm CoVaR}}_{\alpha, \beta}[Y|X]$,
which is the VaR of $Y$ conditional upon $X$ being in a stress scenario,
to $\mbox{\rm VaR}_{\beta}[Y]$, which evaluates $Y$ unconditionally.
Alternatively, one may replace the benchmark $\mbox{\rm VaR}_{\beta}[Y]$
by the conditional VaR of $Y$ given that $X$ exceeds its median \cite[see][]{mainik2014, Adrian2016}.

\begin{Def}\label{defconcovar} Let $\alpha, \beta\in(0,1)$. Then,
\begin{eqnarray*}
\Delta\mbox{{\rm CoVaR}}_{\alpha, \beta}[Y|X]&=&\mbox{{\rm CoVaR}}_{\alpha, \beta}[Y|X]-\mbox{{\rm VaR}}_{\beta}[Y],\\
\Delta^{\rm med}\mbox{{\rm CoVaR}}_{\alpha, \beta}[Y|X]&=&\mbox{{\rm CoVaR}}_{\alpha, \beta}[Y|X]-\mbox{{\rm CoVaR}}_{1/2, \beta}[Y|X].
\end{eqnarray*}
\end{Def}

Risk contribution measures can, of course, also be defined by invoking e.g., CoES, as follows
\citep[see][]{Acharya2010, karimalis2018}:
\begin{Def}\label{defconcoes} Let $\alpha, \beta\in(0,1)$. Then,
\begin{eqnarray*}
\Delta\mbox{{\rm CoES}}_{\alpha, \beta}[Y|X]&=&\mbox{{\rm CoES}}_{\alpha, \beta}[Y|X]-\mbox{{\rm ES}}_{\beta}[Y],\\
\Delta^{\rm med}\mbox{{\rm CoES}}_{\alpha, \beta}[Y|X]&=&\mbox{{\rm CoES}}_{\alpha, \beta}[Y|X]-\mbox{{\rm CoES}}_{1/2, \beta}[Y|X].
\end{eqnarray*}
\end{Def}

\section{Conditional Distortion Risk Measures and Distortion Risk Contribution Measures}\label{sec:distordef}
Consider a bivariate random vector $(X,Y)$ with marginal d.f.'s $F,G\in\mathcal{C}$ and joint d.f. $H\in\mathcal{C}^{2}$.
We define the conditional distortion (CoD) risk measure as a mapping from a bivariate distribution to the real line.
Specifically, the CoD-risk measure is defined as follows.
\begin{Def}\label{defcorisk} For $g, h\in\mathcal{G}$,
\begin{eqnarray*}
\mbox{{\rm CoD}}_{g,h}[Y|X]&=&\mbox{{\rm D}}_{h}\left[Y|X>\mbox{{\rm D}}_{g}[X]\right]\\
&=&-\int_{-\infty}^{0}[1-h(\widebar{F}_{Y|X>{\rm D}_{g}[X]}(y))]\dif y+\int_{0}^{+\infty}h(\widebar{F}_{Y|X>{\rm D}_{g}[X]}(y))\dif y,
\end{eqnarray*}
where $\mbox{{\rm D}}_{g}[X]$ is presented in Definition \ref{defdistor}.
\end{Def}


\begin{Rem}
\begin{itemize}
\item [(a)] Note that in Definition \ref{defcorisk} $g$ and $h$ are distortion functions
imposed on the d.f.'s of $X$ and $[Y|X>\mathrm{D}_{g}[X]]$, respectively.
\item [(b)] Obviously, the CoD-risk measure presented in Definition \ref{defcorisk} contains the CoVaR and CoES as special cases.
More explicitly, we have
\begin{itemize}
\item [(i)] if $g(p)={\bm1}_{(1-\alpha,1]}(p)$ and $h(p)={\bm1}_{(1-\beta,1]}(p)$, then $\mbox{{\rm CoD}}_{g,h}[Y|X]=\mbox{{\rm CoVaR}}_{\alpha, \beta}[Y|X]$;
\item [(ii)] if $g(p)={\bm1}_{(1-\alpha,1]}(p)$ and $h(p)=\min\{1,\frac{p}{1-\beta}\}$, then $\mbox{{\rm CoD}}_{g,h}[Y|X]=\mbox{{\rm CoES}}_{\alpha, \beta}[Y|X]$.
\item [(iii)] if $g(p)={\bm1}_{(1-\alpha,1]}(p)$ and $h(p)=p$, then $\mbox{{\rm CoD}}_{g,h}[Y|X]=\mbox{{\rm MES}}_{\alpha}[Y|X]$.
\end{itemize}
Besides, the CoD-risk measure provides two other related types of conditional risk measures:
\begin{itemize}
\item [(iv)] if $g(p)=\min\{1,\frac{t}{1-\alpha}\}$ and $h(p)={\bm1}_{(1-\beta,1]}(p)$, then $\mbox{{\rm CoD}}_{g,h}[Y|X]=\mbox{\rm VaR}_{\beta}[Y|X>\mbox{\rm ES}_{\alpha}[X]]$;
\item [(v)] if $g(p)=\min\{1,\frac{t}{1-\alpha}\}$ and $h(p)=\min\{1,\frac{p}{1-\beta}\}$, then
\begin{equation*}
\mbox{{\rm CoD}}_{g,h}[Y|X]=\mbox{\rm ES}_{\beta}[Y|X>\mbox{\rm ES}_{\alpha}[X]]=\frac{1}{1-\beta}\int_{\beta}^{1}\mbox{\rm VaR}_{p}[Y|X>\mbox{\rm ES}_{\alpha}[X]]\dif p,
\end{equation*}
which was defined in Equation (10) of \cite{Boyle2012}.
\end{itemize}
\item [(c)] The CoD-risk measure does not in general satisfy the subadditivity property; for example, if $h(p)={\bm1}_{(1-\beta,1]}(p)$, it reduces to the VaR, which is not subadditive in general. However, if $h$ is concave then the CoD-risk measure inherits the subadditivity property of 
    $\mbox{{\rm D}}_{h}[\cdot]$.
\item [(d)] It should be noted that we can replace $\mbox{{\rm D}}_{g}[X]$ by any real value to generalize Definition \ref{defcorisk}. However, all of the existing methods concerned with VaR, median, and ES can be treated as special cases of distortion risk measures under appropriate conditions.
\end{itemize}
\end{Rem}

To illustrate the generality of CoD-risk measures,
the following example provides an illustration of the class of CoD-risk measures
that goes beyond the existing conditional risk measures such as CoVaR, CoES, and MES
that are present in the current literature.
\begin{Exa}\label{exaCoDmax} Assume that $Y$ is a nonnegative r.v. and
consider the distortion function $h(p)=1-(1-p)^k$, for $k\in\mathbb{N}_+$ and $p\in[0,1]$.
Let $\widetilde{Y}^{g}_i$ be independent copies of the conditional r.v. $[Y|X>\mbox{{\rm D}}_{g}[X]]$, for $i=1,\ldots,k$.
According to Definition \ref{defcorisk}, we then have
\begin{eqnarray*}
\mbox{{\rm CoD}}_{g,h}[Y|X]
&=&\int_{0}^{+\infty}h(\widebar{F}_{Y|X>{\rm D}_{g}[X]}(y))\dif y\\
&=& \int_{0}^{+\infty}[1-F_{Y|X>{\rm D}_{g}[X]}^k(y)]\dif y\\
&=&\mathbb{E}[\max\{\widetilde{Y}^{g}_1,\widetilde{Y}^{g}_2,\ldots,\widetilde{Y}^{g}_k\}]\\
&=& \mathbb{E}[\widetilde{Y}^{g}_{k:k}],
\end{eqnarray*}
where $\widetilde{Y}^{g}_{k:k}$ is the maximum order statistic of $\widetilde{Y}^{g}_1,\ldots,\widetilde{Y}^{g}_k$.
This means that the CoD-risk measure can be represented as the expectation of the maximum order statistic
computed from a set of i.i.d. r.v.'s with d.f. $F_{Y|X>{\rm D}_{g}[X]}$.
\end{Exa}

For a given CoD-risk measure,
we can define the associated distortion risk contribution ($\Delta$CoD) measure, as follows.
\begin{Def}\label{defcontririsk} For $g, h\in\mathcal{G}$,
\begin{equation*}
\Delta\mbox{{\rm CoD}}_{g,h}[Y|X]=\mbox{{\rm CoD}}_{g,h}[Y|X]-\mbox{{\rm D}}_{h}[Y],
\end{equation*}
where
\begin{equation*}\label{defg2}
\mbox{{\rm D}}_{h}(Y)=-\int_{-\infty}^{0}[1-h(\widebar{G}(y))]\dif y+\int_{0}^{+\infty}h(\widebar{G}(y))\dif y.
\end{equation*}
\end{Def}

\begin{Rem} The distortion risk contribution measure defined in Definition \ref{defcontririsk} contains $\Delta\mbox{{\rm CoVaR}}$ and $\Delta\mbox{{\rm CoES}}$ as special cases.
Indeed,
\begin{itemize}
\item [(i)] if $g(t)={\bm1}_{(1-\alpha,1]}(p)$ and $h(p)={\bm1}_{(1-\beta,1]}(p)$, then $\Delta\mbox{{\rm CoD}}_{g,h}[Y|X]=\Delta\mbox{{\rm CoVaR}}_{\alpha, \beta}[Y|X]$;
\item [(ii)] if $g(p)={\bm1}_{(1-\alpha,1]}(p)$ and $h(p)=\min\{1,\frac{p}{1-\beta}\}$, then $\Delta\mbox{{\rm CoD}}_{g,h}[Y|X]=\Delta\mbox{{\rm CoES}}_{\alpha, \beta}[Y|X]$.
\end{itemize}
Furthermore, the distortion risk contribution measure also provides two related contribution measures which are absent in the existing literature:
\begin{itemize}
\item [(iii)] if $g(p)=\min\{1,\frac{p}{1-\alpha}\}$ and $h(p)={\bm1}_{(1-\beta,1]}(p)$, then
\begin{equation*}
\Delta\mbox{{\rm CoD}}_{g,h}[Y|X]=\mbox{\rm VaR}_{\beta}[Y|X>\mbox{\rm ES}_{\alpha}[X]]-\mbox{\rm VaR}_{\beta}[Y];
\end{equation*}
\item [(iv)] if $g(p)=\min\{1,\frac{p}{1-\alpha}\}$ and $h(p)=\min\{1,\frac{p}{1-\beta}\}$, then
\begin{eqnarray*}
\Delta\mbox{{\rm CoD}}_{g,h}[Y|X]&=&\mbox{\rm ES}_{\beta}[Y|X>\mbox{\rm ES}_{\alpha}[X]]-\mbox{\rm ES}_{\beta}[Y]\\
&=&\frac{1}{1-\beta}\int_{\beta}^{1}\mbox{\rm VaR}_{p}[Y|X>\mbox{\rm ES}_{\alpha}[X]]\dif p-\frac{1}{1-\beta}\int_{\beta}^{1}\mbox{\rm VaR}_{p}[Y]\dif p.
\end{eqnarray*}
\end{itemize}
It should be mentioned that \cite{Boyle2012} defined one type of risk contribution measure (see their Equation (8)) as
\begin{equation*}
\mbox{\rm ES}_{\beta}[Y|X=\mbox{\rm ES}_{\alpha}[X]]-\mbox{\rm ES}_{\beta}[Y],
\end{equation*}
for which the conditional event is based on $[X=\mbox{\rm ES}_{\alpha}[X]]$ but not $[X>\mbox{\rm ES}_{\alpha}[X]]$.
\end{Rem}

The following example provides the expression of the distortion risk contribution measure of Definition \ref{defcontririsk},
under the setup of Example \ref{exaCoDmax}.
\begin{Exa}\label{exaCoDcont1} Under the setup of Example \ref{exaCoDmax}, we have
\begin{equation*}
\Delta\mbox{{\rm CoD}}_{g,h}[Y|X]=\mathbb{E}[\widetilde{Y}^{g}_{k:k}]-\mathbb{E}\left[Y_{k:k}\right],
\end{equation*}
where $Y_{k:k}$ is the maximum order statistic of $Y_1,\ldots,Y_k$
with $Y_i$ being independent copies of $Y$, for $i=1,\ldots,k$.
\end{Exa}

We can also define risk contribution measures with respect to different distortion functions for the risk $X$,
henceforth sometimes referred to as distortion risk contribution measures of Type-II
to distinguish them from the distortion risk contribution measures of Type-I in Definition \ref{defcontririsk}:
\begin{Def}\label{defconII} For $g, \tilde{g},h\in\mathcal{G}$,
\begin{equation*}
\Delta^{ \tilde{g}}\mbox{{\rm CoD}}_{g,h}[Y|X]=\mbox{{\rm CoD}}_{g,h}[Y|X]-\mbox{{\rm CoD}}_{\tilde{g},h}[Y|X].
\end{equation*}
\end{Def}


\begin{Rem}
It is worth noting that the distortion risk contribution measure of Definition \ref{defconII} contains $\Delta^{\rm med}\mbox{{\rm CoVaR}}$ and $\Delta^{\rm med}\mbox{{\rm CoES}}$ as special cases.
More explicitly,
\begin{itemize}
\item [(i)] if $g(p)={\bm1}_{(1-\alpha,1]}(p)$, $\mbox{{\rm D}}_{g}[X]=F^{-1}(1/2)$, and $h(p)={\bm1}_{(1-\beta,1]}(p)$, then $\Delta^{\tilde{g}}\mbox{{\rm CoD}}_{g,h}[Y|X]=\Delta^{\rm med}\mbox{{\rm CoVaR}}_{\alpha, \beta}[Y|X]$;
\item [(ii)] if $g(p)={\bm1}_{(1-\alpha,1]}(p)$, $\mbox{{\rm D}}_{g}[X]=F^{-1}(1/2)$, and $h(p)=\min\{1,\frac{p}{1-\beta}\}$, then $\Delta^{\tilde{g}}\mbox{{\rm CoD}}_{g,h}[Y|X]=\Delta^{\rm med}\mbox{{\rm CoES}}_{\alpha, \beta}[Y|X]$.
\end{itemize}
\end{Rem}

The next example provides an illustration of the distortion risk contribution measures arising from Definition \ref{defconII}.
\begin{Exa}\label{exaCoDcont2} Under the setup of Example \ref{exaCoDmax},
let $\widetilde{Y}^{\tilde{g}}_i$ be independent copies of the conditional r.v. $[Y|X>\mbox{{\rm D}}_{\tilde{g}}[X]]$, for $i=1,\ldots,k$.
Then,
\begin{equation*}
\Delta\mbox{{\rm CoD}}_{g,h}^{\tilde{g}}[Y|X]=\mathbb{E}[\widetilde{Y}^{g}_{k:k}]-\mathbb{E}[\widetilde{Y}^{\tilde{g}}_{k:k}],
\end{equation*}
where $\widetilde{Y}^{g}_{k:k}$ and $\widetilde{Y}^{\tilde{g}}_{k:k}$
are the maximum order statistics of
$\widetilde{Y}^{g}_1,\ldots,\widetilde{Y}^{g}_k$ and
$\widetilde{Y}^{\tilde{g}}_1,\ldots,\widetilde{Y}^{\tilde{g}}_k$, respectively.
\end{Exa}

In the next theorems, we present some useful expressions and properties of our three types of CoD-risk measures and distortion risk contribution measures introduced in Definitions \ref{defcorisk}, \ref{defcontririsk}, and \ref{defconII}.
\begin{The}\label{theCoDexp} Let $(U,V)\sim C$ where $C$ is a copula of $H$.
If $F$ is continuous and strictly increasing, and $h$ is left-continuous, then
\begin{equation}\label{enewCoD}
\mbox{{\rm CoD}}_{g,h}[Y|X]=\int_{0}^{1}G^{-1}(F^{-1}_{V|U > u_{g}}(p))\dif \widebar{h}(p),
\end{equation}
where $u_{g}=F(\mbox{{\rm D}}_{g}[X])$, $\widebar{h}(p)=1-h(1-p)$ for $p\in[0,1]$, and $F_{V|U > u}(v)=\frac{v-C(u,v)}{1-u}$ for $(u,v)\in(0,1)^2$.
\end{The}
\proof Since $F$ is continuous and strictly increasing, $(U,V)\sim C$, and the marginals of $C$ are uniform, it follows that $\mathbb{P}(U>u_g)=\mathbb{P}(X>\mbox{{\rm D}}_{g}[X])$.
Then, the d.f. of $[Y|X >\mbox{{\rm D}}_{g}[X]]$ can be written as
\begin{eqnarray*}
F_{Y|X >{\rm D}_{g}[X]}(y)&=&\mathbb{P}(Y\leq y|X >\mbox{{\rm D}}_{g}[X])\\
&=&\frac{\mathbb{P}(Y\leq y, X >\mbox{{\rm D}}_{g}[X])}{\mathbb{P}(X >\mbox{{\rm D}}_{g}[X])}\\
&=& \frac{\mathbb{P}(Y\leq y)-\mathbb{P}(Y\leq y, X\leq\mbox{{\rm D}}_{g}[X])}{1-\mathbb{P}(X\leq\mbox{{\rm D}}_{g}[X])}\\
&=& \frac{G(y)-C(F(\mbox{{\rm D}}_{g}[X]),G(y))}{1-F(\mbox{{\rm D}}_{g}[X])}\\
&=& F_{V|U > u_{g}}(G(y)),
\end{eqnarray*}
which in turn implies that $F_{Y|X >{\rm D}_{g}[X]}^{-1}(p)=G^{-1}(F_{V|U > u_{g}}^{-1}(p))$ by using the argument that the event $\{F_{Y|X >{\rm D}_{g}[X]}(y)\geq p\}$ is equivalent to $\{F_{V|U > u_{g}}(G(y))\geq p\}$, for any $p\in(0,1)$.
Hence, by applying Fubini's theorem and a change of variable \citep[see Theorem 6 in][]{dhaene2012remarks} one can verify that
\begin{eqnarray*}
\mbox{{\rm CoD}}_{g,h}[Y|X]&=&-\int_{-\infty}^{0}\left[1-h(1-F_{Y|X >{\rm D}_{g}[X]}(y))\right]\dif y
+\int_{0}^{+\infty}h(1-F_{Y|X >{\rm D}_{g}[X]}(y))\dif y\\
&=& \int_{0}^{1}F_{Y|X >{\rm D}_{g}[X]}^{-1}(1-p)\dif h(p)\\
&=& \int_{0}^{1}F_{Y|X >{\rm D}_{g}[X]}^{-1}(p)\dif \widebar{h}(p)\\
&=& \int_{0}^{1}G^{-1}(F_{V|U> u_{g}}^{-1}(p))\dif \widebar{h}(p).
\end{eqnarray*}
Thus, the proof is established.\qed

\begin{Rem}\label{rem:con} Consider the setup of Theorem \ref{theCoDexp}.
Define the generalized upper inverses $G^{-1+}(p)=\sup\{x\in\mathbb{R}\mid G(x)\leq p\}$ and $F^{-1+}_{V|U > u_{g}}(p)=\sup\{x\in\mathbb{R}\mid F_{V|U > u_{g}}(x)\leq p\}$ with $\sup\emptyset=-\infty$ by convention.
Since the event $\{F_{Y|X>{\rm D}_{g}[X]}(y)\leq p\}$ is equivalent to $\{F_{V|U>u_g}(G(y))\leq p\}$,
we have $F^{-1+}_{Y|X>{\rm D}_{g}[X]}(p)=G^{-1+}(F^{-1+}_{V|U > u_{g}}(p))$, for $p\in(0,1)$.
If now, under the setup of Theorem \ref{theCoDexp}, $h$ were right-continuous instead of left-continuous,
then, by applying Theorem 4 in \cite{dhaene2012remarks}, expression (\ref{enewCoD}) can be modified as
\begin{equation*}\label{enewCoDrc}
\mbox{{\rm CoD}}_{g,h}[Y|X]=\int_{0}^{1}G^{-1+}(F^{-1+}_{V|U > u_{g}}(p))\dif \widebar{h}(p).
\end{equation*}
Note that $F_{V|U > u_{g}}(v)=\frac{v-C(u_g,v)}{1-u_g}$.
If $G$ is continuous and strictly increasing, and $v-C(u,v)$ is continuous and strictly increasing in $v\in[0,1]$ for any $u\in(0,1)$
(which implies that $F_{V|U > u_{g}}$ is continuous and strictly increasing),
we have $G^{-1+}(p)=G^{-1}(p)$ and $F^{-1+}_{V|U > u_{g}}(p)=F^{-1}_{V|U > u_{g}}(p)$, which implies that the distortion function $h$ in Theorem \ref{theCoDexp} can be either left-continuous or right-continuous (given that $F$ is continuous and strictly increasing).
Then, by applying Theorem 7 of \cite{dhaene2012remarks}, $h$ can be also assumed to be any general distortion function,
i.e., a convex combination of left-continuous and right-continuous distortion functions.
\end{Rem}

\begin{Cor} Under the setup of Theorem \ref{theCoDexp},
\begin{itemize}
\item[(i)] if $g(p)={\bm1}_{(1-\alpha,1]}(p)$ and $h(p)={\bm1}_{(1-\beta,1]}(p)$, then
\begin{equation*}
\mbox{{\rm CoD}}_{g,h}[Y|X]=\mbox{{\rm CoVaR}}_{\alpha,\beta}[Y|X]=G^{-1}(F^{-1}_{V|U>\alpha}(\beta)),
\end{equation*}
which is defined in \cite{Girardi2013};
\item[(ii)] if $g(p)={\bm1}_{(1-\alpha,1]}(p)$ and $h(p)=\min\{1,\frac{p}{1-\beta}\}$, then
\begin{equation*}
\mbox{{\rm CoD}}_{g,h}[Y|X]=\mbox{{\rm CoES}}_{\alpha,\beta}[Y|X]=\frac{1}{1-\beta}\int_{\beta}^{1}G^{-1}(F^{-1}_{V|U>\alpha}(p))\dif p,
\end{equation*}
which is defined in \cite{mainik2014}.
\end{itemize}
\end{Cor}

Based on Theorem \ref{theCoDexp}, we have the following result.
\begin{The}\label{theCONexp} Let $(U,V)\sim C$ where $C$ is a copula of $H$. If $F$ is continuous and strictly increasing, and $h$ is left-continuous, then
\begin{equation}\label{econtri1}
\Delta\mbox{{\rm CoD}}_{g,h}[Y|X]=\int_{0}^{1}\left[G^{-1}(F^{-1}_{V|U > u_{g}}(p))-G^{-1}(p)\right]\dif \widebar{h}(p),
\end{equation}
\begin{equation}\label{contrig1}
\Delta^{\tilde{g}}\mbox{{\rm CoD}}_{g,h}[Y|X]=\int_{0}^{1}\left[G^{-1}(F^{-1}_{V|U > u_{g}}(p))-G^{-1}(F^{-1}_{V|U > u_{\tilde{g}}}(p))\right]\dif \widebar{h}(p),
\end{equation}
where $u_{g}=F(\mbox{{\rm D}}_{g}[X])$, $u_{\tilde{g}}=F(\mbox{{\rm D}}_{\tilde{g}}[X])$, $\widebar{h}(p)=1-h(1-p)$ for $p\in[0,1]$, and $F_{V|U > u}(v)=\frac{v-C(u,v)}{1-u}$ for $(u,v)\in(0,1)^2$.
\end{The}

\section{Stochastic Orders and CoD-Risk Measures}\label{sec:diststo}
In the sequel, we always assume that the d.f.'s of $X$ and $X'$ are continuous and strictly increasing
and that the distortion functions for $Y$ and $Y'$ are left-continuous, to avoid unnecessary technical discussions.
We note that, if the d.f.'s of $Y$ and $Y'$ are continuous and strictly increasing and both $v-C(u,v)$ and $v-C'(u,v)$ are continuous and strictly increasing in $v\in[0,1]$ for any $u\in(0,1)$, then, in light of Remark \ref{rem:con}, all of our results can be generalized to the case when the distortion functions for $Y$ and $Y'$ are right-continuous or general, that is, a convex combination of left-continuous and right-continuous distortion functions \citep[see Theorem 7 of][]{dhaene2012remarks}.

\subsection{The Risks $Y$ and $Y'$ Have the Same Distribution}
This subsection considers sufficient conditions for the CoD-risk measures of two bivariate random vectors $(X,Y)$ and $(X',Y')$ to be ordered, 
where $Y$ and $Y'$ have common d.f.'s.
The next theorem states that the CoD-risk measure preserves the ordering 
induced by ``$\prec $'' between the copulas
and by the distortion functions applied to $Y$ and $Y'$.
\begin{The}\label{theGene1} Let $(X,Y)$ and $(X',Y')$ be two bivariate random vectors having the same marginals but different copulas $C$ and $C'$, respectively.
Then, $C\prec C'$ and $h\leq h'$ imply that $\mbox{{\rm CoD}}_{g,h}[Y|X]\leq\mbox{{\rm CoD}}_{g,h'}[Y'|X']$.
\end{The}
\proof Let $(U,V)\sim C$ and $(U',V')\sim C'$.
From Theorem \ref{theCoDexp}, we have
\begin{equation*}
\mbox{{\rm CoD}}_{g,h}[Y|X]=\int_{0}^{1}G^{-1}(F^{-1}_{V|U > u_{g}}(p))\dif \widebar{h}(p),\ \
\mbox{{\rm CoD}}_{g,h'}[Y'|X']=\int_{0}^{1}G^{-1}(F^{-1}_{V'|U' > u_{g}}(p))\dif \widebar{h}'(p),
\end{equation*}
where $\widebar{h}'(p)=1-h'(1-p)$ for $p\in[0,1]$.

We first show that $\mbox{{\rm CoD}}_{g,h}[Y|X]\leq\mbox{{\rm CoD}}_{g,h}[Y'|X']$.
Since $\widebar{h}$ is increasing, this reduces to showing that $G^{-1}(F^{-1}_{V|U > u_{g}}(p))\leq G^{-1}(F^{-1}_{V'|U' > u_{g}}(p))$, i.e., $F^{-1}_{V|U > u_{g}}(p)\leq F^{-1}_{V'|U' > u_{g}}(p)$ for $p\in(0,1)$.
Thus, it suffices to show that $F_{V|U > u_{g}}(t)\geq F_{V'|U' > u_{g}}(t)$ for $t\in(0,1)$, that is,
\begin{equation*}
\frac{t-C(u_{g},t)}{1-u_{g}}\geq\frac{t-C'(u_{g},t)}{1-u_{g}},
\end{equation*}
which is in fact guaranteed by the condition $C\prec C'$.

On the other hand, we can verify that $\widebar{h}(0)=\widebar{h}'(0)=0$, $\widebar{h}(1)=\widebar{h}'(1)=1$ and $\widebar{h}'(p)\leq\widebar{h}(p)$ because of $h(p)\leq h'(p)$ for $p\in[0,1]$.
Then, by using integration by parts, one has
\begin{eqnarray*}
\mbox{{\rm CoD}}_{g,h}[Y'|X']-\mbox{{\rm CoD}}_{g,h'}[Y'|X']&=&\int_{0}^{1}G^{-1}(F^{-1}_{V'|U' > u_{g}}(p))\dif (\widebar{h}(p)-\widebar{h}'(p))\\
&=&\int_{0}^{1}(\widebar{h}'(p)-\widebar{h}(p))\dif G^{-1}(F^{-1}_{V'|U' > u_{g}}(p))\\
&\leq&0,
\end{eqnarray*}
which yields that $\mbox{{\rm CoD}}_{g,h}[Y'|X']\leq\mbox{{\rm CoD}}_{g,h'}[Y'|X']$.
Hence, the proof is established.\qed

The following result, not necessarily requiring $F=F'$, can be easily derived from Theorem \ref{theGene1} when $g$ is the distortion function of VaR.
\begin{Cor}\label{corcovar1} Let $(X,Y)$ and $(X',Y')$ be two bivariate random vectors having copulas $C$ and $C'$, respectively.
Suppose that $G=G'$ and $g(p)={\bm1}_{(1-\alpha,1]}(p)$, for some $\alpha\in(0,1)$.
Then, $C\prec C'$ and $h\leq h'$ imply that $\mbox{{\rm CoD}}_{g,h}[Y|X]\leq\mbox{{\rm CoD}}_{g,h'}[Y'|X']$.
\end{Cor}

\begin{Rem} Theorem \ref{theGene1} and Corollary \ref{corcovar1} do not hold in general 
if $X$ and $X'$ adopt different distortion functions, i.e., if $g\neq g'$.
\end{Rem}

\begin{Rem} Under the additional assumption that $h(p)=h'(p)={\bm1}_{(1-\beta,1]}(p)$, the result of Corollary \ref{corcovar1} reduces to Theorem 3.4 of \cite{mainik2014}.
\end{Rem}

To conclude this subsection, we investigate the effects of threshold quantiles of $X$ and $X'$ and the dependence structure among $(X,Y)$ on the CoD-risk measures.
\begin{The}\label{theg11} Let $(X,Y)$ and $(X',Y')$ be two bivariate random vectors having copulas $C$ and $C'$, respectively.
Suppose that $G=G'$ and $C\prec C'$. Let $u_{g}=F(\mbox{{\rm D}}_{g}[X])$ and $u_{g'}=F'(\mbox{{\rm D}}_{g'}[X'])$.
Then $\mbox{{\rm CoD}}_{g,h}[Y|X]\leq\mbox{{\rm CoD}}_{g',h'}[Y'|X']$ if $h\leq h'$, and either one of the following two conditions holds:
\begin{itemize}
\item [(i)]$u_{g}\leq u_{g'}$ and $Y\uparrow_{\rm RTI}X$ or $Y'\uparrow_{\rm RTI}X'$ or  both hold;
\item [(ii)]$u_{g}\geq u_{g'}$ and $Y\uparrow_{\rm RTD}X$ or $Y'\uparrow_{\rm RTD}X'$ or both hold.
\end{itemize}
\end{The}
\proof We only give the proof for (i).
The proof for (ii) can be established in a similar manner.
We assume that $u_{g}\leq u_{g'}$ and $Y\uparrow_{\rm RTI}X$
(the other two cases follow similarly).
Let $U=F(X)$ and $V=G(Y)$.
In light of Theorem \ref{theCoDexp}, we have
\begin{equation*}
\mbox{{\rm CoD}}_{g,h}[Y|X]=\int_{0}^{1}G^{-1}(F^{-1}_{V|U > u_{g}}(p))\dif \widebar{h}(p).
\end{equation*}
By making use of a change of variable $p=F_{V|U > u_{g}}(t)$, we obtain
\begin{eqnarray*}
\mbox{{\rm CoD}}_{g,h}[Y|X]&=&\int_{0}^{1}G^{-1}(t)\dif \widebar{h}\left(F^{-1}_{V|U > u_{g}}(t)\right)\nonumber\\
&=&\int_{0}^{1}G^{-1}(t)\dif \widebar{h}\left(\frac{t-C(u_{g},t)}{1-u_{g}}\right)\nonumber\\
&=&\int_{0}^{1}G^{-1}(t)\dif \widebar{h}(A_{u_{g}}(t)),
\end{eqnarray*}
where $A_{u_{g}}(t)=1-\frac{\widebar{C}(u_{g},t)}{1-u_{g}}$.
Similarly, by letting $U'=F'(X')$ and $V'=G'(Y')$, we have
\begin{equation*}
\mbox{{\rm CoD}}_{g',h'}[Y'|X']=\int_{0}^{1}G'^{-1}(t)\dif \widebar{h}'(A_{u_{g'}}(t)),
\end{equation*}
where $A_{u_{g'}}(t)=1-\frac{\widebar{C}'(u_{g'},t)}{1-u_{g'}}$. Since $G=G'$, $\widebar{h}'(A_{u_{g'}}(0))=\widebar{h}(A_{u_{g'}}(0))=0$, and $\widebar{h}'(A_{u_{g'}}(1))=\widebar{h}(A_{u_{g'}}(1))=1$, we have
\begin{eqnarray}\label{taske1}
\mbox{{\rm CoD}}_{g',h'}[Y'|X']-\mbox{{\rm CoD}}_{g,h}[Y|X]&=& \int_{0}^{1}G^{-1}(t)\dif \left[\widebar{h}'(A_{u_{g'}}(t))-\widebar{h}(A_{u_{g}}(t))\right]\nonumber\\
&=& \int_{0}^{1}\left[\widebar{h}(A_{u_{g}}(t))-\widebar{h}'(A_{u_{g'}}(t))\right]\dif G^{-1}(t).
\end{eqnarray}
In order to show the nonnegativity of (\ref{taske1}), it suffices to show that $\widebar{h}(A_{u_{g}}(t))\geq\widebar{h}'(A_{u_{g'}}(t))$, for all $t\in[0,1]$.
Since $h\leq h'$, one has $\widebar{h}(A_{u_{g}}(t))\geq\widebar{h}'(A_{u_{g}}(t))$, for all $t\in[0,1]$. Thus, it is enough to show that $\widebar{h}'(A_{u_{g}}(t))\geq\widebar{h}'(A_{u_{g'}}(t))$, for all $t\in[0,1]$, that is,
\begin{equation}\label{taske2}
\frac{\widebar{C}'(u_{g'},t)}{1-u_{g'}}\geq \frac{\widebar{C}(u_{g},t)}{1-u_{g}}.
\end{equation}
Taking into account that $C\prec C'$, we have that
\begin{equation}\label{taske3}
\frac{\widebar{C}'(u_{g'},t)}{1-u_{g'}}\geq \frac{\widebar{C}(u_{g'},t)}{1-u_{g'}}.
\end{equation}
Thus, by using (\ref{taske3}), (\ref{taske2}) can be established if we can show that
\begin{equation}\label{taske4}
\frac{\widebar{C}(u_{g'},t)}{1-u_{g'}}\geq \frac{\widebar{C}(u_{g},t)}{1-u_{g}}.
\end{equation}
Since $u_{g}\leq u_{g'}$ and $Y\uparrow_{\rm RTI}X$, it holds that
\begin{equation*}
\frac{\widebar{C}(u_{g'},t)}{1-u_{g'}}=\mathbb{P}(V>t|U>u_{g'})\geq\mathbb{P}(V>t|U>u_{g})=\frac{\widebar{C}(u_{g},t)}{1-u_{g}},
\end{equation*}
which proves (\ref{taske4}) and thus the desired result is obtained.\qed

Theorem \ref{theg11} states that, if $X$ and $Y$ are positively [negatively] dependent through $Y\uparrow_{\rm RTI~[RTD]}X$, then a larger distortion function employed for risk $Y$, more concordance of the copula, together with a larger [smaller] threshold quantile adopted for risk $X$ lead to a larger CoD-risk measure.

\begin{Rem} Let $(X,Y)$ be a bivariate random vector having copula $C$.
Suppose that $h\leq g$ and $Y\uparrow_{\rm RTD}X$.
Then, in light of Theorem \ref{theg11}(ii), we have $\mbox{{\rm CoD}}_{g,h}[Y|X]\leq\mbox{{\rm CoD}}_{h,g}[Y|X]$.
This means that if $Y$ is negatively dependent of $X$ through RTD, then a larger distortion function for $Y$ and a smaller distortion function for $X$ lead to a larger value of the CoD-risk measure.
\end{Rem}

\begin{Rem} Let $(X,Y)$ and $(X',Y)$ be two bivariate random vectors having the same copula $C$.
Suppose that $h= h'$, and either (i) $u_{g}\leq u_{g'}$ and $Y\uparrow_{\rm RTI}X$ or $Y\uparrow_{\rm RTI}X'$ or both hold, or (ii) $u_{g}\geq u_{g'}$ and  $Y\uparrow_{\rm RTD}X$ or $Y\uparrow_{\rm RTD}X'$ or both hold.
Then Theorem \ref{theg11} implies that $\mbox{{\rm CoD}}_{g,h}[Y|X]\leq\mbox{{\rm CoD}}_{g',h}[Y|X']$ for all $h\in\mathcal{G}$.
This result states that $X'$ is more relevant for $Y$ than $X$ if $Y$ is positively [negatively] dependent of $X$ (and/or $X'$) through RTI [RTD]
and the threshold quantile of $X$ is smaller [larger] than that of $X'$,
which is consistent with the systemic relevance order proposed in Definition 12 of \cite{DLZorder2018a}.
\end{Rem}

\subsection{The Risks $Y$ and $Y'$ Have Different Distributions}
In this subsection, we present sufficient conditions on the dependence structure and distortion functions for the CoD-risk measures of two bivariate random vectors $(X,Y)$ and $(X',Y')$ to be ordered when $Y$ and $Y'$ have different d.f.'s.

\cite{sordo2007} provided a useful characterization of the usual stochastic order and the increasing convex order as follows.
In a similar manner, we can give an equivalent characterization of the increasing concave order.
\begin{Lem}\label{lemmaicx} Let $X$ and $Y$ be two r.v.'s with d.f.'s $F$ and $G$, respectively.
Then, $X\leq_{\rm st~[icx, ~icv]}Y$ if and only if
\begin{equation*}
\int_{0}^{1}F^{-1}(t)\dif \phi(t)\leq\int_{0}^{1}G^{-1}(t)\dif \phi(t),
\end{equation*}
for all increasing [increasing convex, increasing concave] $\phi:[0,1]\to[0,1]$.
\end{Lem}
\proof The proof for the usual stochastic order and the increasing convex order can be found in \cite{sordo2007}.
We only prove the characterization of the increasing concave order.

Assume that $X\leq_{\rm icv}Y$.
According to Theorem 4.A.1 in \cite{shaked2007sto}, we know that $X\leq_{\rm icv}Y$ is equivalent to $-X\geq_{\rm icx}-Y$.
Then, by using the equivalent characterization of the increasing convex order, it follows that
\begin{equation*}
-X\geq_{\rm icx}-Y\Longleftrightarrow\int_{0}^{1}F_{-X}^{-1}(t)\dif \phi(t)\geq\int_{0}^{1}F_{-Y}^{-1}(t)\dif \phi(t),
\end{equation*}
for all increasing convex $\phi:[0,1]\to[0,1]$.
Note that
\begin{equation*}
\int_{0}^{1}F_{-X}^{-1}(t)\dif \phi(t)=-\int_{0}^{1}F^{-1}(1-t)\dif \phi(t).
\end{equation*}
Hence, we have
\begin{equation*}
X\leq_{\rm icv}Y\Longleftrightarrow-X\geq_{\rm icx}-Y\Longleftrightarrow\int_{0}^{1}F^{-1}(1-t)\dif \phi(t)\leq\int_{0}^{1}G^{-1}(1-t)\dif \phi(t),
\end{equation*}
that is
\begin{equation*}
X\leq_{\rm icv}Y\Longleftrightarrow\int_{0}^{1}F^{-1}(t)\dif \psi(t)\leq\int_{0}^{1}G^{-1}(t)\dif \psi(t),
\end{equation*}
where $\psi(t):=1-\phi(1-t)$ is increasing and concave on $[0,1]$.
Hence, the proof is established. \qed

Next, for two given random vectors $(X,Y)$ and $(X',Y')$, we present sufficient conditions 
in terms of stochastic orders of the marginal d.f.'s of $Y$ and $Y'$, 
the respective copulas and dependence structure, and the distortion functions
for their CoD-risk measures to be ordered.
\begin{The}\label{theicxco1} Let $(X,Y)$ and $(X',Y')$ be two bivariate random vectors having copulas $C$ and $C'$, respectively.
Suppose that $F=F'$, $C\prec C'$, and $h\leq h'$.
\begin{itemize}
\item [(i)] Suppose that $X\uparrow_{\rm SI}Y$ or $X'\uparrow_{\rm SI}Y'$ or both hold. 
Then, $Y\leq_{\rm st~[icx]}Y'$ implies that $\mbox{{\rm CoD}}_{g,h}[Y|X]\leq\mbox{{\rm CoD}}_{g,h'}[Y'|X']$ for any $g\in\mathcal{G}$ and increasing [increasing concave] $h,h'\in\mathcal{G}$.
\item [(ii)] Suppose that $X\uparrow_{\rm SD}Y$ or $X'\uparrow_{\rm SD}Y'$ or both hold. 
Then, $Y\leq_{\rm icv}Y'$ implies that $\mbox{{\rm CoD}}_{g,h}[Y|X]\leq\mbox{{\rm CoD}}_{g,h'}[Y'|X']$ for any $g\in\mathcal{G}$ and increasing convex $h,h'\in\mathcal{G}$.
\end{itemize}
\end{The}
\proof We only give the proof for the increasing convex ordering between $Y$ and $Y'$.
The proofs for the usual stochastic ordering and the increasing concave ordering can be obtained in a similar manner by using Lemma \ref{lemmaicx}.
Furthermore, we only consider the case of $X\uparrow_{\rm SI}Y$ since the proof can be carried out similarly for $X'\uparrow_{\rm SI}Y'$.

Let $U=F(X)$ and $V=G(Y)$. In light of Theorem \ref{theCoDexp}, we have
\begin{equation*}
\mbox{{\rm CoD}}_{g,h}[Y|X]=\int_{0}^{1}G^{-1}(F^{-1}_{V|U > u_{g}}(p))\dif \widebar{h}(p).
\end{equation*}
By making use of a change of variable $p=F_{V|U > u_{g}}(t)$, we obtain
\begin{equation}\label{chava1}
\mbox{{\rm CoD}}_{g,h}[Y|X]=\int_{0}^{1}G^{-1}(t)\dif \widebar{h}(A(t)),
\end{equation}
where $A(t)=1-\frac{\widebar{C}(u_{g},t)}{1-u_{g}}$. 
Note that $A(t)$ is increasing and convex in $t\in[0,1]$ since $\frac{\dif A(t)}{\dif t}\stackrel{\rm sgn}{=}-\frac{\partial\widebar{C}(u_{g},t)}{\partial t}=\mathbb{P}(U>u_{g}|V=t)$ is nonnegative and increasing in $t\in[0,1]$ because $X\uparrow_{\rm SI}Y$. 
On the other hand, it is easy to verify that $\widebar{h}$ is also increasing convex due to the increasing concavity of $h$. Hence, we know $\widebar{h}(A(t))$ is increasing and convex in $t\in[0,1]$. Similarly, by $F=F'$ we can obtain
\begin{equation*}
\mbox{{\rm CoD}}_{g,h'}[Y|X]=\int_{0}^{1}G'^{-1}(t)\dif \widebar{h}'(B(t)),
\end{equation*}
where $B(t)=1-\frac{\widebar{C}'(u_{g},t)}{1-u_{g}}$. The desired result boils down to showing that
\begin{equation}\label{icxe0}
\int_{0}^{1}G^{-1}(t)\dif \widebar{h}(A(t))\leq\int_{0}^{1}G'^{-1}(t)\dif \widebar{h}'(B(t)).
\end{equation}

On the one hand, by using Lemma \ref{lemmaicx}, $Y\leq_{\rm icx}Y'$ implies that
\begin{equation}\label{icxe1}
\int_{0}^{1}G^{-1}(t)\dif \widebar{h}(A(t))\leq\int_{0}^{1}G'^{-1}(t)\dif \widebar{h}(A(t)).
\end{equation}
On the other hand, $C\prec C'$ implies that $A(t)\geq B(t)$, and thus $\widebar{h}(A(t))\geq \widebar{h}(B(t))\geq \widebar{h}'(B(t))$ since $h\leq h'$. Because $\widebar{h}(A(0))=\widebar{h}'(B(0))=0$ and $\widebar{h}(A(1))=\widebar{h}'(B(1))=1$, we then have
\begin{equation*}
\int_{0}^{1}G'^{-1}(t)\dif \left[\widebar{h}(A(t))-\widebar{h}'(B(t))\right]=\int_{0}^{1}\left[\widebar{h}'(B(t))-\widebar{h}(A(t))\right]\dif G'^{-1}(t)\leq0,
\end{equation*}
which means that
\begin{equation}\label{icxe2}
\int_{0}^{1}G'^{-1}(t)\dif \widebar{h}(A(t))\leq\int_{0}^{1}G'^{-1}(t)\dif \widebar{h}'(B(t)).
\end{equation}
Upon combining (\ref{icxe1}) and (\ref{icxe2}), the desired result (\ref{icxe0}) is established.\qed

\begin{Rem} Definition 2 of \cite{DLZorder2018a} defines the systemic contribution order 
in terms of the increasing convex order.
More explicitly, consider the market of losses $\underline{Z}$, the corresponding microprudential regulation $\underline{R}$, and the aggregate residual loss level $s\in\mathbb{R}_{\geq 0}$.
Individual loss $Z_{j}$ is said to be ``smaller in systemic contribution order'' (denoted by $Z_{j}\leq_{(\underline{R},s)-con}Z_{k}$) than individual loss $Z_{k}$ under microprudential regulation $\underline{R}$ and aggregate loss level $s$,
if
\begin{equation*}
\left[\left(Z_{j}-R_{j}\right)_{+}\mid \sum_{i=1}^{n}\left(Z_{i}-R_{i}\right)_{+}>s\right]
\leq_{\rm icx}\left[\left(Z_{k}-R_{k}\right)_{+}\mid \sum_{i=1}^{n}\left(Z_{i}-R_{i}\right)_{+}>s\right].
\end{equation*}
According to Theorem \ref{theicxco1}(i), if $X=X'=\sum_{i=1}^{n}\left(Z_{i}-R_{i}\right)_{+}$, $Y=(Z_{j}-R_{j})_{+}$, $Y'=(Z_{k}-R_{k})_{+}$, $C=C'$, $h=h'$, and $\sum_{i=1}^{n}\left(Z_{i}-R_{i}\right)_{+}\uparrow_{\rm SI}(Z_{j}-R_{j})_{+}$ or $\sum_{i=1}^{n}\left(Z_{i}-R_{i}\right)_{+}\uparrow_{\rm SI}(Z_{k}-R_{k})_{+}$ or both hold, then
$\left(Z_{j}-R_{j}\right)_{+}\leq_{\rm icx}\left(Z_{k}-R_{k}\right)_{+}$ implies that
\begin{equation*}
\mbox{{\rm CoD}}_{g,h}\left[(Z_{j}-R_{j})_{+}\big|\sum_{i=1}^{n}\left(Z_{i}-R_{i}\right)_{+}\right]\leq\mbox{{\rm CoD}}_{g,h}\left[(Z_{k}-R_{k})_{+}\big|\sum_{i=1}^{n}\left(Z_{i}-R_{i}\right)_{+}\right],
\end{equation*}
for all concave $h$, which is consistent with the definition $Z_{j}\leq_{(\underline{R},s)-con}Z_{k}$
when taking $s=\mbox{{\rm D}}_{g}[\sum_{i=1}^{n}\left(Z_{i}-R_{i}\right)_{+}]$.
\end{Rem}

The following result, not necessarily requiring $F=F'$, can be derived from Theorem \ref{theicxco1} when $g$ corresponds to the distortion function of VaR.
\begin{Cor}\label{coricxco1} Let $(X,Y)$ and $(X',Y')$ be two bivariate random vectors having copulas $C$ and $C'$, respectively.
Suppose that $g(p)={\bm1}_{(1-\alpha,1]}(p)$, $C\prec C'$, and $h\leq h'$.
\begin{itemize}
\item [(i)]Suppose that $X\uparrow_{\rm SI}Y$ or $X'\uparrow_{\rm SI}Y'$ or both hold.
    Then, $Y\leq_{\rm st~[icx]}Y'$ implies that $\mbox{{\rm CoD}}_{g,h}[Y|X]\leq\mbox{{\rm CoD}}_{g,h'}[Y'|X']$ for any increasing [increasing concave] $h,h'\in\mathcal{G}$.
\item [(ii)]Suppose that $X\uparrow_{\rm SD}Y$ or $X'\uparrow_{\rm SD}Y'$ or both hold.
    Then, $Y\leq_{\rm icv}Y'$ implies that $\mbox{{\rm CoD}}_{g,h}[Y|X]\leq\mbox{{\rm CoD}}_{g,h'}[Y'|X']$ for any increasing convex $h,h'\in\mathcal{G}$.
\end{itemize}
\end{Cor}
\begin{Rem} In the special case that $h(p)=h'(p)=\min\{1,\frac{p}{1-\beta}\}$,
the result of Corollary \ref{coricxco1}(i) reduces to Theorem 12 in \cite{sordo2018IME}.
\end{Rem}

The next result generalizes Theorem \ref{theicxco1} to the case of different d.f.'s of $X$ and $X'$, 
where we can replace the condition `$F=F'$' by requiring that $F(\mbox{{\rm D}}_{g}[X])\leq F'(\mbox{{\rm D}}_{g'}[X'])$.
\begin{The}\label{DIFFgene1} Let $(X,Y)$ and $(X',Y')$ be two bivariate random vectors having copulas $C$ and $C'$, respectively.
Suppose that $C\prec C'$, $h\leq h'$, and $u_{g}\leq u_{g'}$, where $u_{g}=F(\mbox{{\rm D}}_{g}[X])$ and $u_{g'}=F'(\mbox{{\rm D}}_{g'}[X'])$.
\begin{itemize}
\item [(i)] Suppose that $(X,Y)$ or $(X',Y')$ or both are PDS. 
Then, $Y\leq_{\rm st~[icx]}Y'$ implies that $\mbox{{\rm CoD}}_{g,h}[Y|X]\leq\mbox{{\rm CoD}}_{g',h'}[Y'|X']$ for any increasing [increasing concave] $h,h'\in\mathcal{G}$.
\item [(ii)] Suppose that $(X,Y)$ or $(X',Y')$ or both are NDS. 
Then, $Y\leq_{\rm icv}Y'$ implies that $\mbox{{\rm CoD}}_{g,h}[Y|X]\leq\mbox{{\rm CoD}}_{g',h'}[Y'|X']$ for any increasing convex $h,h'\in\mathcal{G}$.
\end{itemize}
\end{The}
\proof Note that if $(X,Y)$ is PDS [NDS], then $X\uparrow_{\rm SI~[SD]}Y$ and $Y\uparrow_{\rm SI~[SD]}X$.
The proof is then easily obtained by combining the proof methods in Theorems \ref{theg11} and \ref{theicxco1}.\qed

\section{Stochastic Orders and Distortion Risk Contribution Measures}\label{sec:contridis}
Consider two bivariate random vectors $(X,Y)$ and $(X',Y')$. 
This section provides sufficient conditions in terms of stochastic orders of the marginal d.f.'s of $Y$ and $Y'$, 
the respective copulas and dependence structure, and the distortion functions
for their distortion risk contribution measures to be ordered.

\subsection{Dispersive Order and Distortion Risk Contribution Measures}
The following lemma, adapted from \cite{sordo2018IME}, is helpful to establish our main results
linking the dispersive order between marginals with distortion risk contribution measures.
\begin{Lem}\label{lemsddisp} Let $X$ and $Y$ be two continuous r.v.'s with d.f.'s $F$ and $G$, respectively. 
Let $h$ be a convex distortion function and let $g$ be another right-continuous distortion function such
that $h (p)\geq g (p)$, for all $p\in[0, 1]$.
Denote by $X_h$ [$Y_g$] the distorted r.v.'s induced from $X$ [$Y$] by the distortion functions $h$ [$g$].
If $X\leq_{\rm disp}Y$, then
\begin{itemize}
\item [(i)] $F^{-1}_{X_{h}}(p)-F^{-1}_{X}(p)\geq F^{-1}_{Y_{g}}(p)-G^{-1}(p)$, for $p\in(0,1)$;
\item [(ii)] $F^{-1}_{X_{h}}(p)-F^{-1}_{X_g}(p)\geq F^{-1}_{Y_{h}}(p)-F^{-1}_{Y_g}(p)$, for $p\in(0,1)$.
\end{itemize}
\end{Lem}
\proof The proof can be obtained by using similar arguments as in Lemma 14 in \cite{sordo2018IME},
and thus is omitted here for brevity.\qed

\subsubsection{Type-I Distortion Risk Contribution Measures: $\Delta\mbox{{\rm CoD}}_{g,h}[Y|X]$}\label{subsubtypeI}
We first study the sufficient conditions for $\Delta\mbox{{\rm CoD}}_{g,h}[Y|X]$.
\begin{The}\label{thedispCON1} Let $(X,Y)$ and $(X',Y')$ be two bivariate random vectors having copulas $C$ and $C'$, respectively.
Suppose that $F=F'$ and $Y\leq_{\rm disp}Y'$.
\begin{itemize}
\item [(i)] If $C\prec C'$, and $X\uparrow_{\rm SI}Y$ or $X'\uparrow_{\rm SI}Y'$ or both hold, then $\Delta\mbox{{\rm CoD}}_{g,h}[Y|X]\leq\Delta\mbox{{\rm CoD}}_{g,h}[Y'|X']$ for any $g,h\in\mathcal{G}$.
\item [(ii)] If $C\succ C'$, and $X\uparrow_{\rm SD}Y$ or $X'\uparrow_{\rm SD}Y'$ or both hold, then $\Delta\mbox{{\rm CoD}}_{g,h}[Y|X]\geq\Delta\mbox{{\rm CoD}}_{g,h}[Y'|X']$ for any $g,h\in\mathcal{G}$.
\end{itemize}
\end{The}
\proof We only give the proof of (i) since the proof of (ii) is similar by applying Lemma \ref{lemsddisp}(i).
Suppose that $X\uparrow_{\rm SI}Y$.
According to Theorem \ref{theCONexp} and the proof of Theorem \ref{theCoDexp}, we have
\begin{align*}
\Delta\mbox{{\rm CoD}}_{g,h}[Y|X]&=\int_{0}^{1}\left[F^{-1}_{Y_{h}}(p)-G^{-1}(p)\right]\dif \widebar{h}(p),\\
\Delta\mbox{{\rm CoD}}_{g,h}[Y'|X']&=\int_{0}^{1}\left[F^{-1}_{Y'_{h'}}(p)-G'^{-1}(p)\right]\dif \widebar{h}(p),
\end{align*}
where $Y_h=[Y|X >\mbox{{\rm D}}_{g}[X]]$ and $Y'_{h'}=[Y'|X' >\mbox{{\rm D}}_{g}[X]]$ are the distorted r.v.'s induced from $Y$ and $Y'$ by the concave distortion functions
\begin{equation}\label{disthh2}
h(p)=\frac{\widebar{C}(F(\mbox{{\rm D}}_{g}[X]),1-p)}{1-F(\mbox{{\rm D}}_{g}[X])}\quad\mbox{and}\quad h'(p)=\frac{\widebar{C}'(F(\mbox{{\rm D}}_{g}[X]),1-p)}{1-F(\mbox{{\rm D}}_{g}[X])},\quad\mbox{$p\in[0,1]$}.
\end{equation}
Since $C\prec C'$, it clearly holds that $h(p)\leq h'(p)$ for all $p\in[0,1]$. From $Y\leq_{\rm disp}Y'$ and Lemma 14 in  \cite{sordo2018IME}, we have
\begin{equation*}
F^{-1}_{Y_{h}}(p)-G^{-1}(p)\leq F^{-1}_{Y'_{h'}}(p)-G'^{-1}(p),\quad\mbox{for $p\in(0,1)$},
\end{equation*}
which yields the desired result since $\widebar{h}(p)$ is increasing in $p\in[0,1]$.\qed

The next theorem generalizes Theorem \ref{thedispCON1} to the case where $X$ and $X'$ may have different d.f.'s.
\begin{The}\label{thedispCON8} Let $(X,Y)$ and $(X',Y')$ be two bivariate random vectors having copulas $C$ and $C'$, respectively.
Let $u_{g}=F(\mbox{{\rm D}}_{g}[X])$ and $u_{g'}=F'(\mbox{{\rm D}}_{g'}[X'])$.
Suppose that $Y\leq_{\rm disp}Y'$.
\begin{itemize}
\item [(i)] If $C$ is PDS, $u_{g}\leq u_{g'}$, and $C\prec C'$, then $\Delta\mbox{{\rm CoD}}_{g,h}[Y|X]\leq\Delta\mbox{{\rm CoD}}_{g',h}[Y'|X']$ for any $h\in\mathcal{G}$.
\item [(ii)] If $C$ is NDS, $u_{g}\geq u_{g'}$, and $C\succ C'$, then $\Delta\mbox{{\rm CoD}}_{g,h}[Y|X]\geq\Delta\mbox{{\rm CoD}}_{g',h}[Y'|X']$ for any $h\in\mathcal{G}$.
\end{itemize}
\end{The}

The following result is partially taken from Theorem 5 in \cite{sordo2015IME} and the proof for the case of a convex distortion function can be established using similar arguments as in the proof of Theorem 5 in \cite{sordo2015IME}.
\begin{Lem}\label{conhr} Let $X$ be a r.v. and let $g$ be a concave [convex] distortion function. Then $X\leq_{\rm hr}[\geq_{\rm hr}]X_g$, where $X_g$ is the distorted r.v. induced from $X$ by applying the distortion function $g$.
\end{Lem}

Recall that a (nonnegative) r.v. $X$ has an increasing [decreasing] failure rate (IFR [DFR])
if, and only if, its survival function $\widebar{F}$ is log-concave [log-convex].

\begin{The}\label{theDFRCo} Let $(X,Y)$ be a bivariate random vector with copula $C$.
Assume that $Y$ is DFR.
Then $\Delta\mbox{{\rm CoD}}_{g,h}[Y|X]\leq\Delta\mbox{{\rm CoD}}_{g,h'}[Y|X]$ for any $g\in\mathcal{G}$ if either one of the following two conditions holds:
\begin{itemize}
\item [(i)]$X\uparrow_{\rm SI}Y$ and $h\leq h'$;
\item [(ii)]$X\uparrow_{\rm SD}Y$ and $h\geq h'$.
\end{itemize}
\end{The}
\proof We only give the proof for (i) since the proof can be established in a similar manner for (ii).
According to Theorem \ref{theCONexp} and the proof of Theorem \ref{theCoDexp}, we have
\begin{equation*}
\Delta\mbox{{\rm CoD}}_{g,h}[Y|X]=\int_{0}^{1}\left[F^{-1}_{Y_{\hat{h}}}(p)-G^{-1}(p)\right]\dif \widebar{h}(p),
\end{equation*}
where $Y_{\hat{h}}=[Y|X >\mbox{{\rm D}}_{g}[X]]$ is a distorted r.v. induced from $Y$ by the concave distortion function
\begin{equation*}
\hat{h}(p)=\frac{\widebar{C}(F(\mbox{{\rm D}}_{g}[X]),1-p)}{1-F(\mbox{{\rm D}}_{g}[X])},\quad\mbox{$p\in[0,1]$}.
\end{equation*}
Then, from Lemma \ref{conhr}, we have $Y\leq_{\rm hr}Y_{\hat{h}}$.
Since $Y$ is DFR, it follows that $Y\leq_{\rm disp}Y_{\hat{h}}$ upon invoking Theorem 3.B.20(a) of \cite{shaked2007sto}.
Therefore, it holds that
\begin{equation*}
G^{-1}(p_2)-G^{-1}(p_1)\leq F^{-1}_{Y_{\hat{h}}}(p_2)-F^{-1}_{Y_{\hat{h}}}(p_1),\quad\mbox{for $0<p_1<p_2<1$},
\end{equation*}
which implies that $F^{-1}_{Y_{\hat{h}}}(p)-G^{-1}(p)$ is increasing in $p\in(0,1)$.
Since $\widebar{h}(0)=\widebar{h}'(0)=0$ and $\widebar{h}(1)=\widebar{h}'(1)=1$, we then have that
\begin{eqnarray*}
\Delta\mbox{{\rm CoD}}_{g,h}[Y|X]-\Delta\mbox{{\rm CoD}}_{g,h'}[Y|X]&=&\int_{0}^{1}\left[F^{-1}_{Y_{\hat{h}}}(p)-G^{-1}(p)\right]\dif [\widebar{h}(p)-\widebar{h}'(p)]\\
&=&\int_{0}^{1}[\widebar{h}'(p)-\widebar{h}(p)]\dif\left[F^{-1}_{Y_{\hat{h}}}(p)-G^{-1}(p)\right]\leq0,
\end{eqnarray*}
which yields the desired result.\qed

\begin{Rem} If $X\uparrow_{\rm SI}Y$, $Y$ is DFR, $g(p)={\bm1}_{(1-\alpha,1]}(p)$, $h(p)=\min\{1,\frac{p}{1-\beta}\}$, and $h'(p)=\min\{1,\frac{p}{1-\beta'}\}$ such that $\beta\leq\beta'$, then Theorem \ref{theDFRCo} reduces to the result of Theorem 17 in \cite{sordo2018IME}.
\end{Rem}

Upon combining Theorems \ref{thedispCON1} and \ref{theDFRCo}, the following result can be obtained immediately, which generalizes the result of Corollary 19 in \cite{sordo2018IME}.
\begin{Cor} Let $(X,Y)$ and $(X',Y')$ be two bivariate random vectors having copulas $C$ and $C'$, respectively.
Suppose that $F=F'$, $X\uparrow_{\rm SI}Y$ or $X'\uparrow_{\rm SI}Y'$ or both hold, and $Y$ or $Y'$ or both are DFR.
Then, $Y\leq_{\rm disp}Y'$, $C\prec C'$, and $h\leq h'$ imply that $\Delta\mbox{{\rm CoD}}_{g,h}[Y|X]\leq \Delta\mbox{{\rm CoD}}_{g,h'}[Y'|X']$ for any $g\in\mathcal{G}$.
\end{Cor}

The following result generalizes the above result to the case of risks $X$ and $X'$ having different d.f.'s.
\begin{The}\label{thetotaldisp} Let $(X,Y)$ and $(X',Y')$ be two bivariate random vectors having copulas $C$ and $C'$, respectively.
Let $u_{g}=F(\mbox{{\rm D}}_{g}[X])$ and $u_{g'}=F'(\mbox{{\rm D}}_{g'}[X'])$.
Suppose that $(X, Y)$ or $(X', Y')$ or both are PDS, and  $Y$ or $Y'$ or both are DFR.
Then, $Y\leq_{\rm disp}Y'$, $C\prec C'$, $u_{g}\leq u_{g'}$, and $h\leq h'$ imply that $\Delta\mbox{{\rm CoD}}_{g,h}[Y|X]\leq \Delta\mbox{{\rm CoD}}_{g',h'}[Y'|X']$ for any $g\in\mathcal{G}$.
\end{The}
\proof By using the proof method in Theorem \ref{DIFFgene1}, the result can be obtained from Theorem \ref{thedispCON8}. \qed

\subsubsection{Type-II Distortion Risk Contribution Measures: $\Delta^{\tilde{g}}\mbox{{\rm CoD}}_{g,h}[Y|X]$}\label{subsubtypeII}

In this subsection, we turn our attention to studying how the dependence structure, threshold quantile of $X$, and the stochastic ordering according to which $Y$ varies change the value of the distortion risk contribution measure $\Delta^{\tilde{g}}\mbox{{\rm CoD}}_{g,h}[Y|X]$.
\begin{The}\label{medthedisp1} Let $(X,Y)$ and $(X',Y')$ be two bivariate random vectors having the same copula $C$.
Suppose that $F=F'$ and $u_{g}\geq u_{\tilde{g}}$, where $u_{g}=F(\mbox{{\rm D}}_{g}[X])$ and $u_{\tilde{g}}=F(\mbox{{\rm D}}_{\tilde{g}}[X])$.
Then, $Y\leq_{\rm disp}Y'$ and $C$ is PDS imply that $\Delta^{\tilde{g}}\mbox{{\rm CoD}}_{g,h}[Y|X]\leq\Delta^{\tilde{g}}\mbox{{\rm CoD}}_{g,h}[Y'|X']$ for any $h\in\mathcal{G}$.
\end{The}
\proof In light of Theorem \ref{theCONexp} and $F=F'$, one can observe that
\begin{equation*}
\Delta^{\tilde{g}}\mbox{{\rm CoD}}_{g,h}[Y|X]=\int_{0}^{1}\left[F^{-1}_{Y_{\hat{h}_{1}}}(p)-F^{-1}_{Y_{\hat{h}_{2}}}(p)\right]\dif \widebar{h}(p),
\end{equation*}
\begin{equation*}
\Delta^{\tilde{g}}\mbox{{\rm CoD}}_{g,h}[Y'|X']=\int_{0}^{1}\left[F^{-1}_{Y'_{\hat{h}_{1}}}(p)-F^{-1}_{Y'_{\hat{h}_{2}}}(p)\right]\dif \widebar{h}(p),
\end{equation*}
where $Y_{\hat{h}_{1}}=[Y|X >\mbox{{\rm D}}_{g}[X]]$ and $Y_{\hat{h}_{2}}=[Y|X>\mbox{{\rm D}}_{\tilde{g}}[X]]$ are the distorted r.v.'s induced from $Y$ by the concave distortion function (this is due to the fact that $C$ is PDS implies that $Y\uparrow_{\rm SI}X$)
\begin{equation}\label{dispdise4}
\hat{h}_{1}(p)=\frac{\widebar{C}(u_{g},1-p)}{1-u_{g}}\quad\mbox{and}\quad \hat{h}_{2}(p)=\frac{\widebar{C}(u_{\tilde{g}},1-p)}{1-u_{\tilde{g}}},\quad\mbox{$p\in[0,1]$},
\end{equation}
and $Y'_{\hat{h}_{1}}=[Y'|X' >\mbox{{\rm D}}_{g}[X']]$ and $Y'_{\hat{h}_{2}}=[Y'|X'>\mbox{{\rm D}}_{\tilde{g}}[X']]$ 
are also the distorted r.v.'s induced from $Y'$ by (\ref{dispdise4}).

On the other hand, the condition that $C$ is PDS implies that $V$ is right tail increasing in $U$ if $(U,V)\sim C$.
Thus, we know
\begin{equation*}
\frac{\widebar{C}(u,1-p)}{1-u}=\mathbb{P}(V>1-p| U>u)
\end{equation*}
is increasing in $u\in[0,1)$ for $p\in[0,1]$.
Therefore, it follows that $\hat{h}_{1}(p)\geq \hat{h}_{2}(p)$ for $p\in[0,1]$ because of $u_{g}\geq u_{\tilde{g}}$.
Then, the desired result can be obtained from Lemma 14 in \cite{sordo2018IME}.\qed

\begin{Rem} Theorem \ref{medthedisp1} contains Theorem 20 of \cite{sordo2018IME} as a special case when $\tilde{g}(p)={\bm1}_{(1/2,1]}(p)$, $g(p)={\bm1}_{(1-\alpha,1]}(p)$, and $h(p)={\bm1}_{(1-\beta,1]}(p)$ with $1/2\leq\alpha\leq1$.
It is also worth noting that the condition that ``$C$ is TP$_2$'' used there can be weakened by ``$C$ is PDS'' as seen in Theorem \ref{medthedisp1}.
Besides, the condition $u_{g}\geq u_{\tilde{g}}$ is equivalent to $\mbox{{\rm D}}_{g}[X]\geq\mbox{{\rm D}}_{\tilde{g}}[X]$.
Therefore, a sufficient condition for this is to require $g\geq \tilde{g}$.
\end{Rem}

The next result provides some other sufficient conditions in terms of the negative dependence structure of the copula.
\begin{The}\label{medthedisp2} Let $(X,Y)$ and $(X',Y')$ be two bivariate random vectors having the same copula $C$.
Suppose that $F=F'$, $u_{g}=F(\mbox{{\rm D}}_{g}[X])$, and $u_{\tilde{g}}=F(\mbox{{\rm D}}_{\tilde{g}}[X])$.
Then, $Y\leq_{\rm disp}Y'$, $C$ is NDS and $u_{g}\leq u_{\tilde{g}}$ imply that $\Delta^{\tilde{g}}\mbox{{\rm CoD}}_{g,h}[Y|X]\geq\Delta^{\tilde{g}}\mbox{{\rm CoD}}_{g,h}[Y'|X']$ for any $h\in\mathcal{G}$.
\end{The}
\proof Upon using Lemma \ref{lemsddisp}(ii), the proof can be established in a similar manner to that of Theorem \ref{medthedisp1} and is thus omitted here for brevity.\qed

\subsection{Excess Wealth Order and Distortion Risk Contribution Measures}


For a r.v. $X$ with d.f. $F$, \cite{sordo2008IME} established an equivalence characterization between the excess wealth order and the class of risk measures of the form
\begin{equation*}
\mbox{{\rm D}}_{\phi_1,\phi_2}[X]=\int_{0}^1F^{-1}(t)\dif \phi_1(t)-\int_{0}^1F^{-1}(t)\dif \phi_2(t),
\end{equation*}
where $\phi_1$ and $\phi_2$ are two distortion functions.
\begin{Lem}{\rm \citep{sordo2008IME}}\label{lemmaew} Let $X$ and $Y$ be two r.v.'s with d.f.'s $F$ and $G$, respectively. 
Then, $X\leq_{\rm ew}Y$ if and only if $\mbox{{\rm D}}_{\phi_1,\phi_2}[X]\leq\mbox{{\rm D}}_{\phi_1,\phi_2}[Y]$ for all $\mbox{{\rm D}}_{\phi_1,\phi_2}$ such that $\phi_2(t)$ and $\phi_1\phi_2^{-1}(t)$ are convex on $t\in[0,1]$.
\end{Lem}
\begin{The}\label{theewCON1} Let $(X,Y)$ and $(X',Y')$ be two bivariate random vectors having copulas $C$ and $C'$, respectively.
Suppose that $F=F'$, $X\uparrow_{\rm SI}Y$ or $X'\uparrow_{\rm SI}Y'$ or both hold, $h(t)$ is concave, and $\widebar{h}(A(\widebar{h}^{-1}(t)))$ is convex, where $A(t)=1-\frac{\widebar{C}(u_{g},t)}{1-u_{g}}$.
Then, $Y\leq_{\rm ew}Y'$ and $C\prec C'$ imply that $\Delta\mbox{{\rm CoD}}_{g,h}[Y|X]\leq \Delta\mbox{{\rm CoD}}_{g,h}[Y'|X']$ for any $g\in\mathcal{G}$.
\end{The}
\proof Assume that $X\uparrow_{\rm SI}Y$ (the case $X'\uparrow_{\rm SI}Y'$ can be dealt with analogously).
Note that
\begin{equation*}
\Delta\mbox{{\rm CoD}}_{g,h}[Y|X]=\int_{0}^{1}G^{-1}(t)\dif \widebar{h}(A(t))-\int_{0}^{1}G^{-1}(t)\dif \widebar{h}(t),
\end{equation*}
\begin{equation*}
\Delta\mbox{{\rm CoD}}_{g,h}[Y'|X']=\int_{0}^{1}G'^{-1}(t)\dif \widebar{h}(B(t))-\int_{0}^{1}G'^{-1}(t)\dif \widebar{h}(t).
\end{equation*}
Since $h(t)$ is concave and $\widebar{h}(A(\widebar{h}^{-1}(t)))$ is convex, it follows from Lemma \ref{lemmaew} that
\begin{eqnarray*}
\Delta\mbox{{\rm CoD}}_{g,h}[Y|X]&\leq&\int_{0}^{1}G'^{-1}(t)\dif \widebar{h}(A(t))-\int_{0}^{1}G'^{-1}(t)\dif \widebar{h}(t)\\
&\leq&\int_{0}^{1}G'^{-1}(t)\dif \widebar{h}(B(t))-\int_{0}^{1}G'^{-1}(t)\dif \widebar{h}(t)\\
&=&\Delta\mbox{{\rm CoD}}_{g,h'}[Y'|X'],
\end{eqnarray*}
where the last inequality is due to the fact that $C\prec C'$ implies $\int_{0}^{1}G'^{-1}(t)\dif \widebar{h}(A(t))\leq\int_{0}^{1}G'^{-1}(t)\dif \widebar{h}(B(t))$.
Hence, the proof is established.\qed

It is interesting to also study sufficient conditions for the ordering of $\Delta\mbox{{\rm CoD}}_{g,h}^{\tilde{g}}[Y|X]$ and $\Delta\mbox{{\rm CoD}}_{g,h}^{\tilde{g}}[Y'|X']$ by using the excess wealth order among the marginals. 
This is left as an open problem.


\section{Interaction between Paired Risks under CoD-Risk Measures and Distortion Risk Contribution Measures}\label{sec:paired}
Recently, \cite{fang2018RISKS} studied how the marginal d.f.'s and the dependence structure affect the interactions among paired risks
under the CoVaR, CoES, $\Delta$CoVaR, and $\Delta$CoES measures.
In this section, we shall establish some novel results for our CoD-risk measures and distortion risk contribution measures, which generalize the corresponding ones established in \cite{fang2018RISKS}.

\begin{The}\label{thepair1} Let $(X,Y)$ be a bivariate random vector with copula $C$.
Assume that $C(u,v)$ is symmetric, $u_{g}^{X}=F(\mbox{{\rm D}}_{g}[X])$, and $u_{g}^{Y}=G(\mbox{{\rm D}}_{g}[Y])$.
\begin{itemize}
\item [(i)]If $X\leq_{\rm st}Y$, $u_{g}^{X}\geq u_{g}^{Y}$, and $Y\uparrow_{\rm RTI}X$, we have $\mbox{{\rm CoD}}_{g,h}[X|Y]\leq \mbox{{\rm CoD}}_{g,h}[Y|X]$ for any $h\in\mathcal{G}$.
\item [(ii)]If $X\leq_{\rm st}Y$, $u_{g}^{X}\leq u_{g}^{Y}$, and $Y\uparrow_{\rm RTD}X$, we have $\mbox{{\rm CoD}}_{g,h}[X|Y]\leq \mbox{{\rm CoD}}_{g,h}[Y|X]$ for any $h\in\mathcal{G}$.
\item [(iii)]If $X\leq_{\rm icx}Y$, $u_{g}^{X}\geq u_{g}^{Y}$, and $C$ is PDS, we have $\mbox{{\rm CoD}}_{g,h}[X|Y]\leq \mbox{{\rm CoD}}_{g,h}[Y|X]$ for any concave $h\in\mathcal{G}$.
\item [(iv)]If $X\leq_{\rm icv}Y$, $u_{g}^{X}\leq u_{g}^{Y}$, and $C$ is NDS,  we have $\mbox{{\rm CoD}}_{g,h}[X|Y]\leq \mbox{{\rm CoD}}_{g,h}[Y|X]$ for any convex $h\in\mathcal{G}$.
\item [(v)]If $X\leq_{\rm disp}Y$, $u_{g}^{X}\geq u_{g}^{Y}$, and $C$ is PDS, we have $\Delta\mbox{{\rm CoD}}_{g,h}[X|Y]\leq \Delta\mbox{{\rm CoD}}_{g,h}[Y|X]$ for any $h\in\mathcal{G}$.
\item [(vi)]If $X\leq_{\rm disp}Y$, $u_{g}^{X}\leq u_{g}^{Y}$, and $C$ is NDS, we have $\Delta\mbox{{\rm CoD}}_{g,h}[X|Y]\geq \Delta\mbox{{\rm CoD}}_{g,h}[Y|X]$ for any $h\in\mathcal{G}$.
\end{itemize}
\end{The}
\proof \underline{\emph{Proof of (i) and (ii)}}: By using (\ref{chava1}), the desired result is equivalent to showing that
\begin{equation*}
\int_{0}^{1}F^{-1}(t)\dif \widebar{h}(\tilde{A}(t))\leq\int_{0}^{1}G^{-1}(t)\dif \widebar{h}(A(t)),
\end{equation*}
where $A(t)=1-\frac{\widebar{C}(u_{g}^{X},t)}{1-u_{g}^{X}}$ and $\tilde{A}(t)=1-\frac{\widebar{C}(t,u_{g}^{Y})}{1-u_{g}^{Y}}$.
Since $C(u,v)$ is symmetric, we have $\tilde{A}(t)=1-\frac{\widebar{C}(u_{g}^{Y},t)}{1-u_{g}^{Y}}$.
By using $X\leq_{\rm st}Y$, we have
\begin{equation*}
\int_{0}^{1}F^{-1}(t)\dif \widebar{h}(\tilde{A}(t))\leq\int_{0}^{1}G^{-1}(t)\dif \widebar{h}(\tilde{A}(t)).
\end{equation*}
On the other hand, in light of $u_{g}^{X}\geq[\leq] u_{g}^{Y}$ and $Y\uparrow_{\rm RTI~[RTD]}X$, one can verify that $A(t)\leq \tilde{A}(t)$.
Thus, it holds that
\begin{equation*}
\int_{0}^{1}G^{-1}(t)\dif \widebar{h}(\tilde{A}(t))-\int_{0}^{1}G^{-1}(t)\dif \widebar{h}(A(t))=\int_{0}^{1}[\widebar{h}(A(t))-\widebar{h}(\tilde{A}(t))]\dif G^{-1}(t)\leq0.
\end{equation*}
Hence, the proof is completed.

\underline{\emph{Proof of (iii) and (iv)}}: In light of the proof of (i) and (ii) and the proof of Theorem \ref{theicxco1},
it is easy to see that both $\tilde{A}(t)$ and $A(t)$ are increasing and convex due to $C$ being PDS.
Besides, $A(t)\leq \tilde{A}(t)$ due to $u_{g}^{X}\geq u_{g}^{Y}$ and $C$ being PDS.
Thus, the concavity of $h$ implies that $\widebar{h}(A(t))$ and $\widebar{h}(\tilde{A}(t))$ are increasing and convex.
Then, the proof of (iii) is completed by using Lemma \ref{lemmaicx} and the second part of the proof of (i) and (ii).
Result (iv) can be proved in a similar manner and thus is omitted here.

\underline{\emph{Proof of (v) and (vi)}}: The proof can be obtained by using that of (i) and (ii), Theorem \ref{thedispCON1}, and Theorem \ref{thedispCON8}.\qed

The next result can be proved easily by using similar arguments as in the proof of Theorem \ref{theewCON1}, and thus we omit the proof for brevity.
\begin{The}\label{thepair2} Let $(X,Y)$ be a bivariate random vector with copula $C$.
Assume that $C(u,v)$ is symmetric and $u_{g}^{X}\geq u_{g}^{Y}$.
If $X\leq_{\rm ew}Y$, $C(u,v)$ is PDS, $h$ is concave, and $\widebar{h}(A_X(\widebar{h}^{-1}(t)))$ is convex, where $A_X(t)=1-\frac{\widebar{C}(u_{g}^{X},t)}{1-u_{g}^{X}}$, then $\Delta\mbox{{\rm CoD}}_{g,h}[X|Y]\leq \Delta\mbox{{\rm CoD}}_{g,h}[Y'|X']$ for any $g\in\mathcal{G}$.
\end{The}

It would be of great interest to obtain sufficient conditions for ordering co-risk measures and risk contribution measures when the copula is asymmetric. This research question is left as an open problem.

\section{Numerical Examples}\label{sec:nume}
This section provides some numerical examples to illustrate our main findings.
Based on our results developed in the previous sections, the choice of the d.f.'s and distortion functions of $X$ and $X'$ can be arbitrary since we only need the relation between $u_{g}$ and $u_{g'}$/$u_{\tilde{g}}$.
Therefore, we do not specify the explicit d.f.'s of $X$ and $X'$ in most of our examples.
We shall provide illustrations of our main results both for positive and negative dependence structures,
which are represented by the Gumbel copula and the Farlie-Gumbel-Morgenstern (FGM) copula, respectively.

\begin{figure}[htbp!]
	\centering
	\subfigure[]{\includegraphics[width=.48\textwidth, height=0.3\textheight]{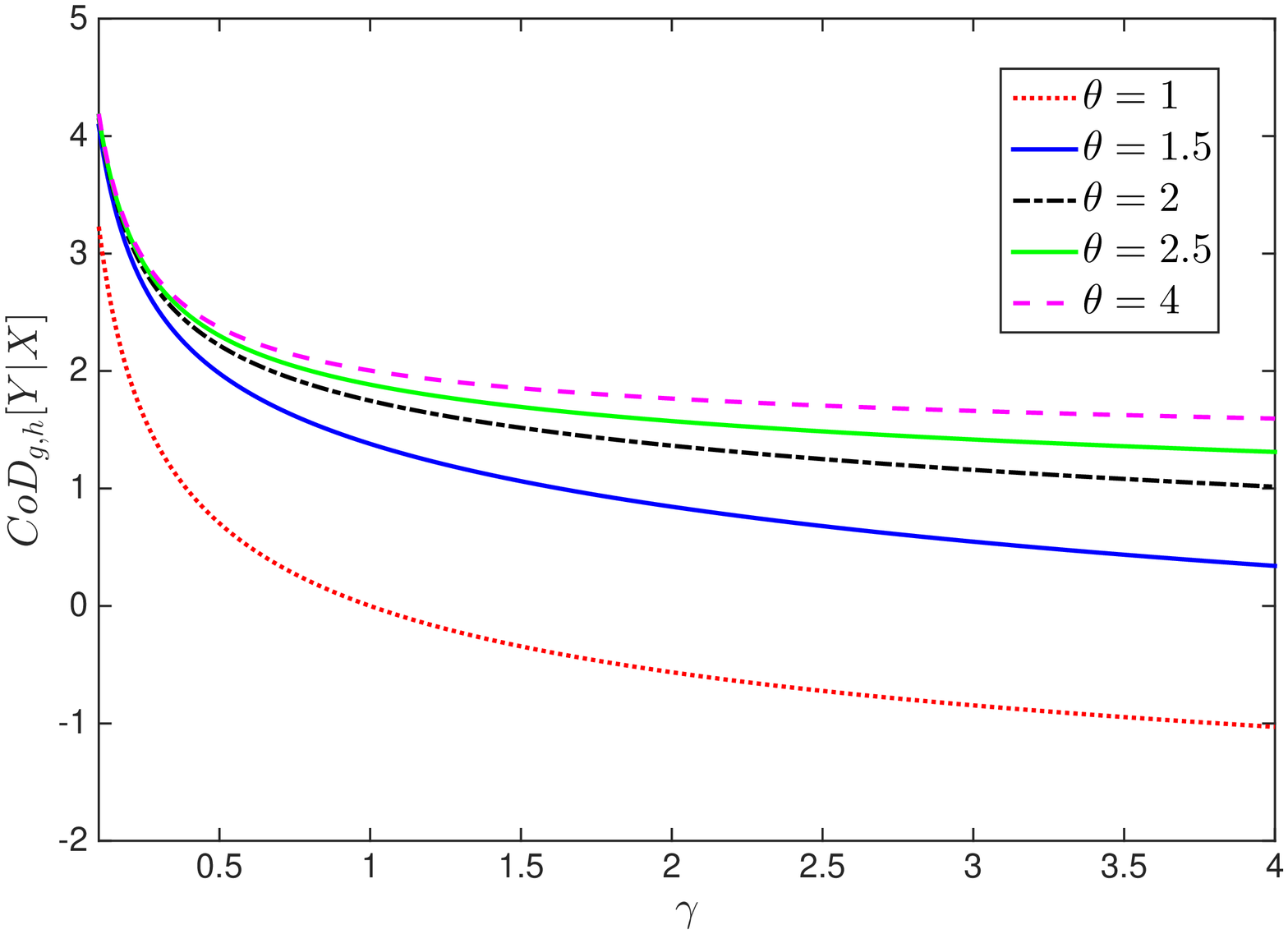}\label{figcod1}}
	\subfigure[]{\includegraphics[width=.48\textwidth, height= 0.3\textheight]{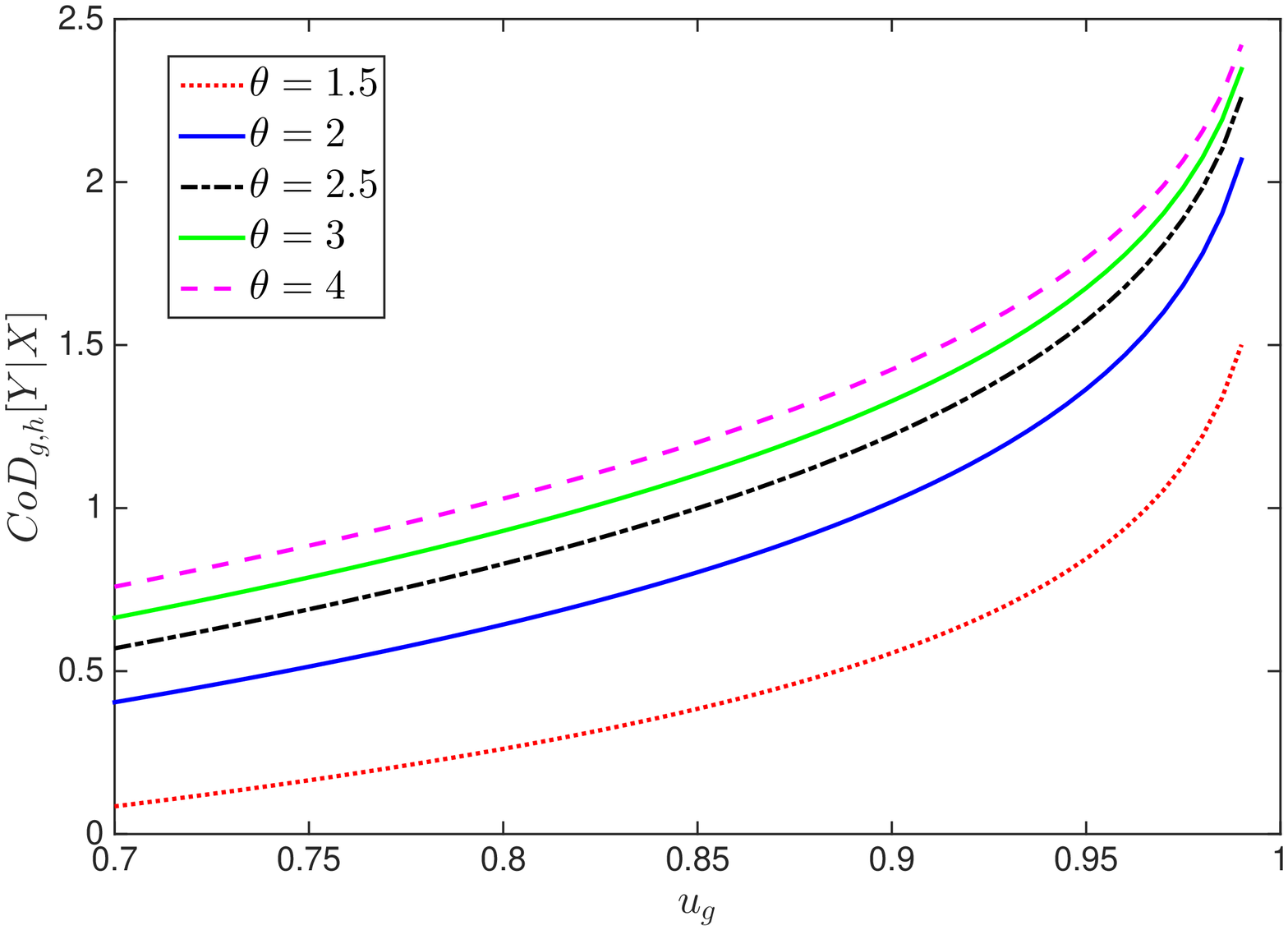}\label{figcod2}}
	\subfigure[]{\includegraphics[width=.48\textwidth, height= 0.3\textheight]{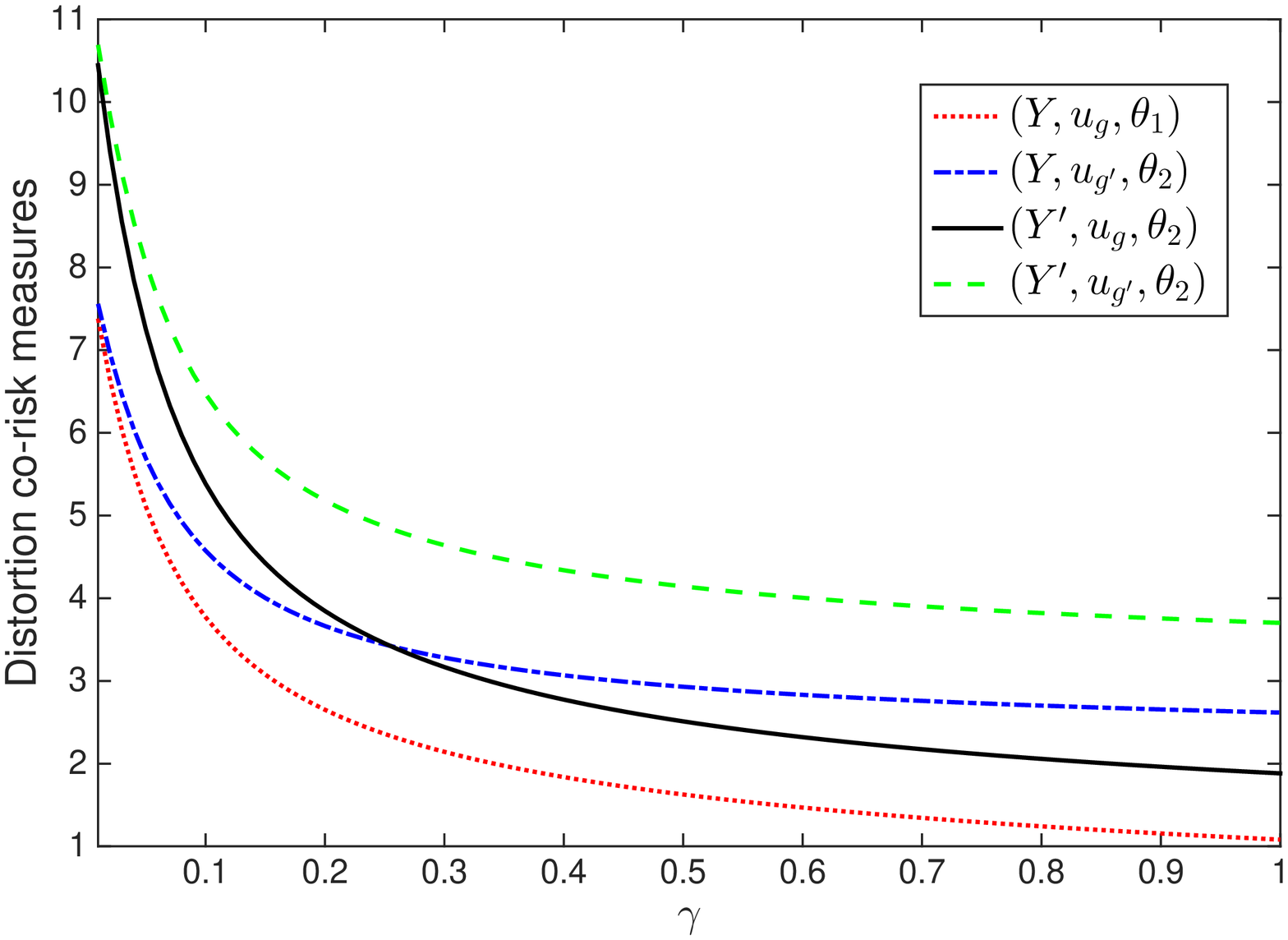}\label{figcod3}}
	\caption{\footnotesize(a) Plot of $\mbox{{\rm CoD}}_{g,h}[Y|X]$ on $\gamma\in[0.1,4]$ for different values of $\theta$. (b) Plot of $\mbox{{\rm CoD}}_{g,h}[Y|X]$ on $u_{g}\in[0.7,0.99]$ for different values of $\theta$. (c) Plot of CoD-risk measures on $\gamma\in(0,1]$ under different settings of d.f., threshold quantile, and dependence parameter.}\label{figcod}
\end{figure}

\subsection{The Gumbel Copula}
The Gumbel copula is defined as
\begin{equation*}
C_{\theta}(u,v)=\exp\left(-\left((-\log u)^{\theta}+(-\log v)^{\theta}\right)^{1/\theta}\right), \quad\mbox{$\theta\geq1$}.
\end{equation*}
It corresponds to the independence copula when $\theta=1$, and to the comonotonic copula when $\theta=\infty$.
It can be inferred from \cite{wei2002} that $C_{\theta}\prec C_{\theta'}$ if $\theta\leq\theta'$.
Besides, $C_{\theta}$ is PDS for all $\theta\geq1$.
Interested readers are referred to \cite{Joe1997} and \cite{Nelsen2007} for more discussions.

\begin{Exa}[CoD-risk measures]
Assume that $Y$ has a standard normal d.f. and $u_{g}=0.95$ for some chosen d.f. of $X$ and distortion function $g$.
Let $h(p)=p^{\gamma}$ for $\gamma>0$.
Note that $h(p)$ is decreasing in $\gamma$ for any $p\in[0,1]$.
\begin{itemize}
\item [(a)] For different values of the dependence parameter $\theta=1,1.5,2,2.5,4$, we plot the values of $\mbox{{\rm CoD}}_{g,h}[Y|X]$ for $\gamma>0$ in Figure \ref{figcod1}.
    It is readily apparent that the CoD-risk measure decreases as the distortion function of $Y$ gets smaller (i.e., $\gamma$ gets larger) for fixed dependence parameter $\theta$, and it increases when the positive dependence gets stronger (i.e., $\theta$ gets larger).
    This illustrates the result of Theorem \ref{theGene1}.
\item [(b)] For different values of the dependence parameter $\theta=1.5,2,2.5,3,4$, we plot the values of $\mbox{{\rm CoD}}_{g,h}[Y|X]$ as $u_{g}$ varies from 0.7 to 0.99 in Figure \ref{figcod2}, from which we observe that the CoD-risk measure increases as the threshold quantile $u_{g}$ gets larger for fixed dependence parameter $\theta$, and it increases when the positive dependence gets stronger (i.e., $\theta$ gets larger).
    Therefore, the theoretical finding in Theorem \ref{theg11}(i) is verified.
\item [(c)] Consider $Y\sim N(0,1)$ and $Y'\sim N(0,2)$ such that $Y\leq_{\rm icx}Y'$ but $Y\nleq_{\rm st}Y'$.
    Assume that $\theta_1=2$, $\theta_2=4$, $u_{g}=0.8$, and $u_{g'}=0.99$.
    Figure \ref{figcod3} gives the plots of $\mbox{{\rm CoD}}_{g,h}[Y|X]$, $\mbox{{\rm CoD}}_{g',h}[Y|X']$, $\mbox{{\rm CoD}}_{g,h}[Y'|X]$, and $\mbox{{\rm CoD}}_{g',h}[Y'|X']$ for different values of $\gamma\in(0,1]$, which implies that $h(p)$ is increasing and concave on $p\in[0,1]$.
    It is readily apparent that these four types of CoD-risk measures become smaller as $\gamma$ increases,
    i.e., as the distortion function becomes smaller.
    Moreover, for any fixed $\gamma\in(0,1]$, we have
\begin{equation*}
\mbox{{\rm CoD}}_{g,h}[Y|X]\leq\mbox{{\rm CoD}}_{g',h}[Y|X']\leq\mbox{{\rm CoD}}_{g',h}[Y'|X'],
\end{equation*}
\begin{equation*}
\mbox{{\rm CoD}}_{g,h}[Y|X]\leq \mbox{{\rm CoD}}_{g,h}[Y'|X]\leq\mbox{{\rm CoD}}_{g',h}[Y'|X'],
\end{equation*}
while $\mbox{{\rm CoD}}_{g',h}[Y|X']$ and $\mbox{{\rm CoD}}_{g,h}[Y'|X]$ cannot be compared.
These observations validate the results of Theorem \ref{DIFFgene1}(i).
\end{itemize}
\end{Exa}

\begin{figure}[!htbp]
	\centering
	\subfigure[]{\includegraphics[width=.48\textwidth, height=0.3\textheight]{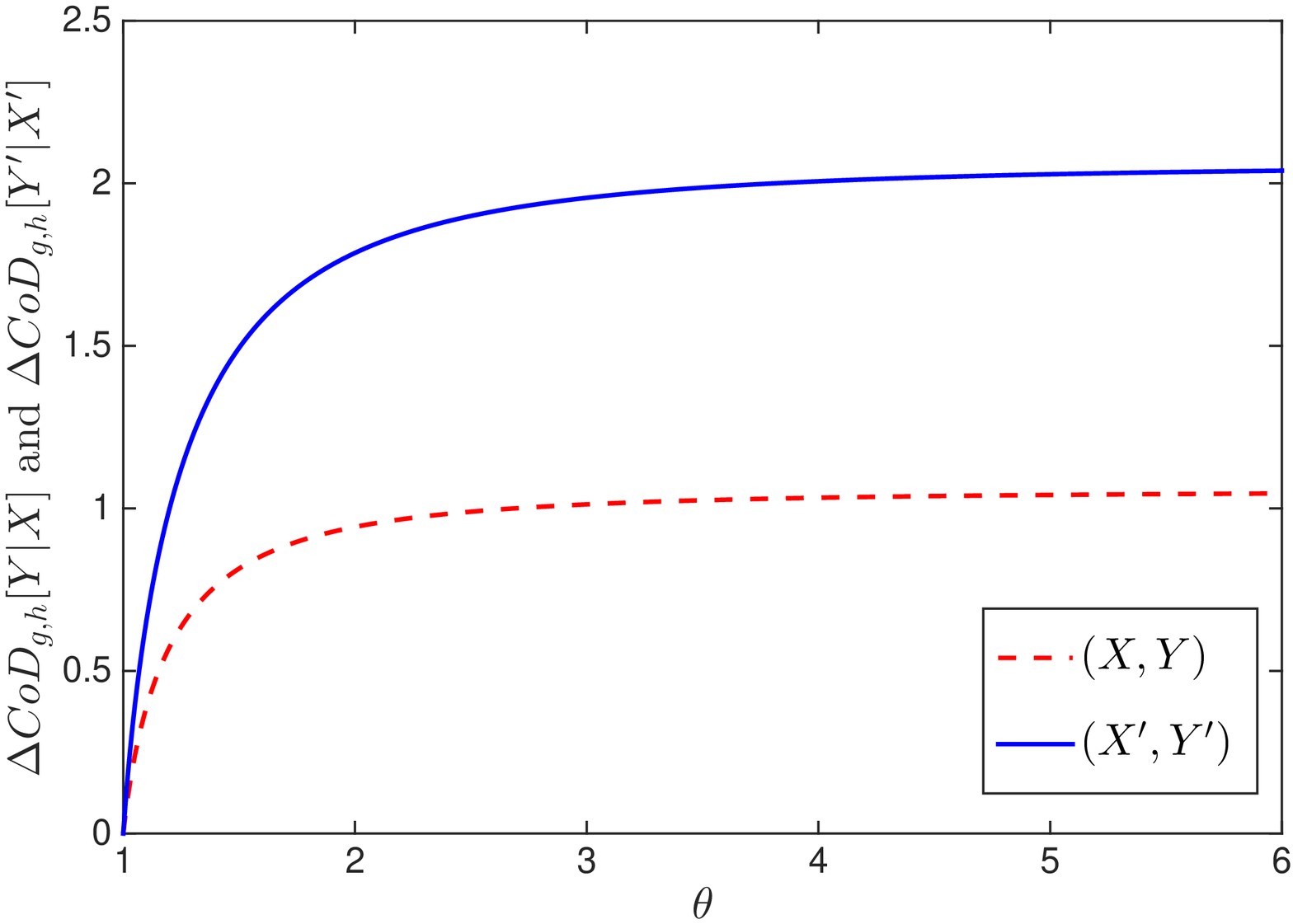}\label{figcontri1}}
	\subfigure[]{\includegraphics[width=.48\textwidth, height= 0.3\textheight]{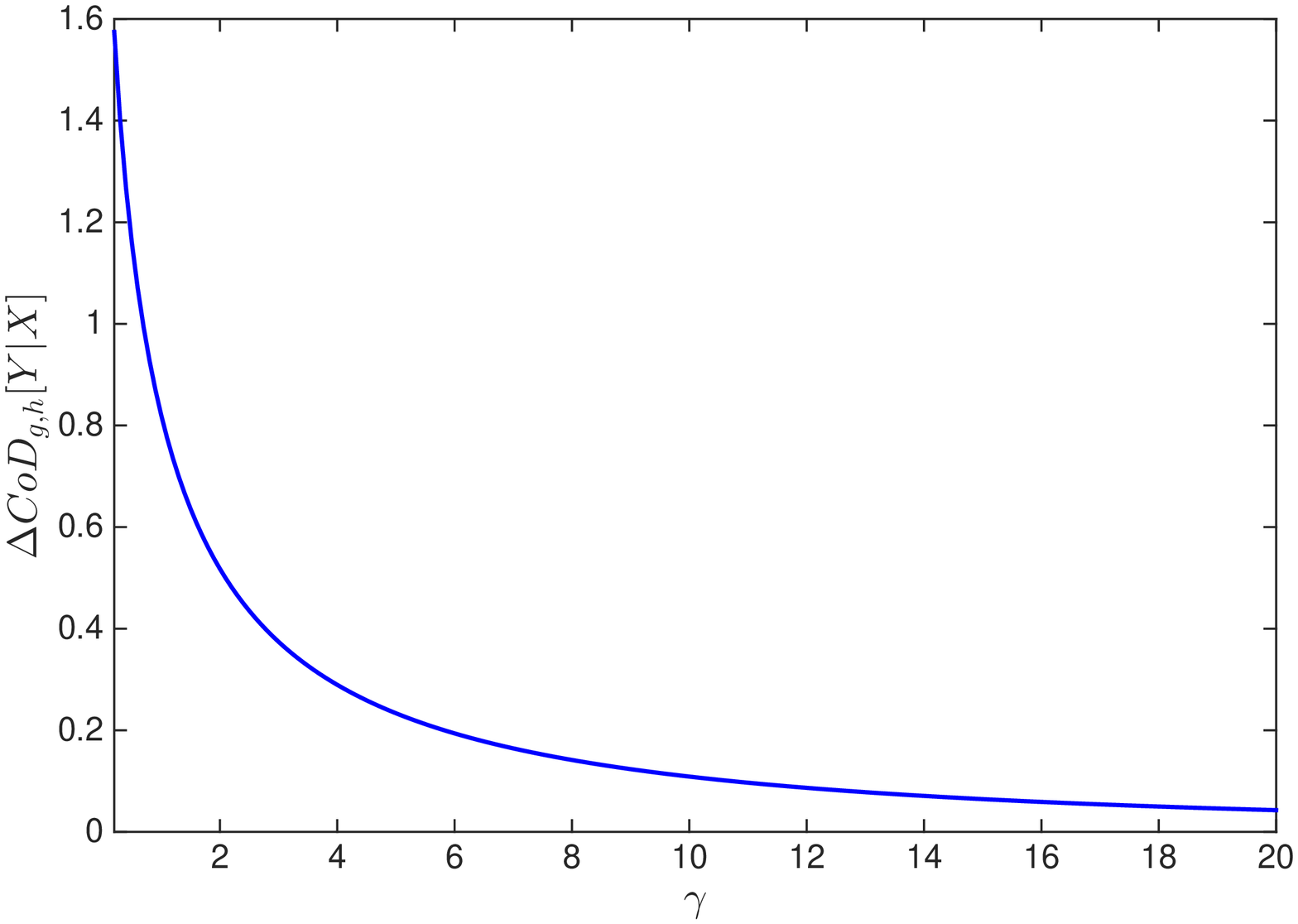}\label{figcontri2}}
	\subfigure[]{\includegraphics[width=.48\textwidth, height= 0.3\textheight]{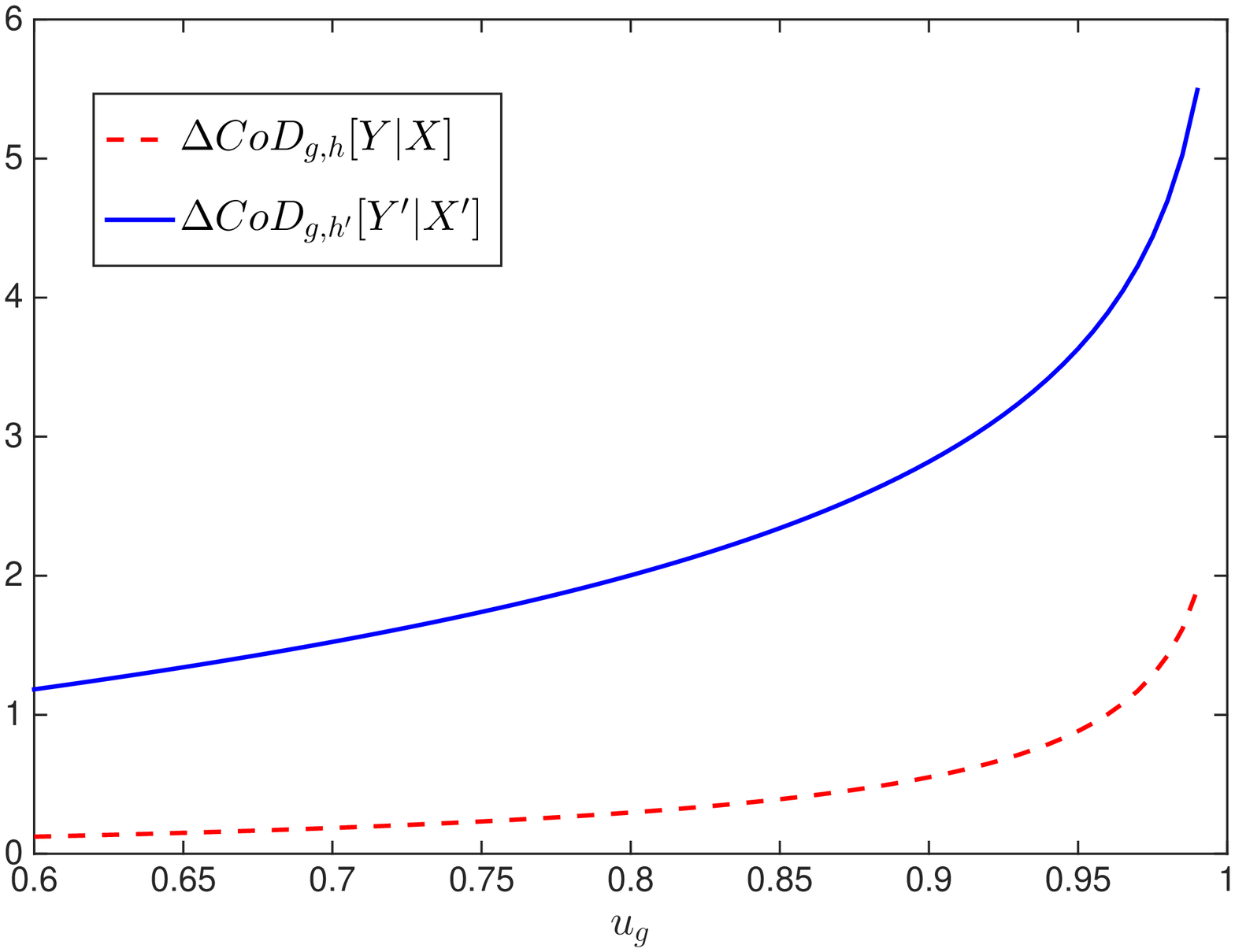}\label{figcontri3}}
	\subfigure[]{\includegraphics[width=.48\textwidth, height= 0.3\textheight]{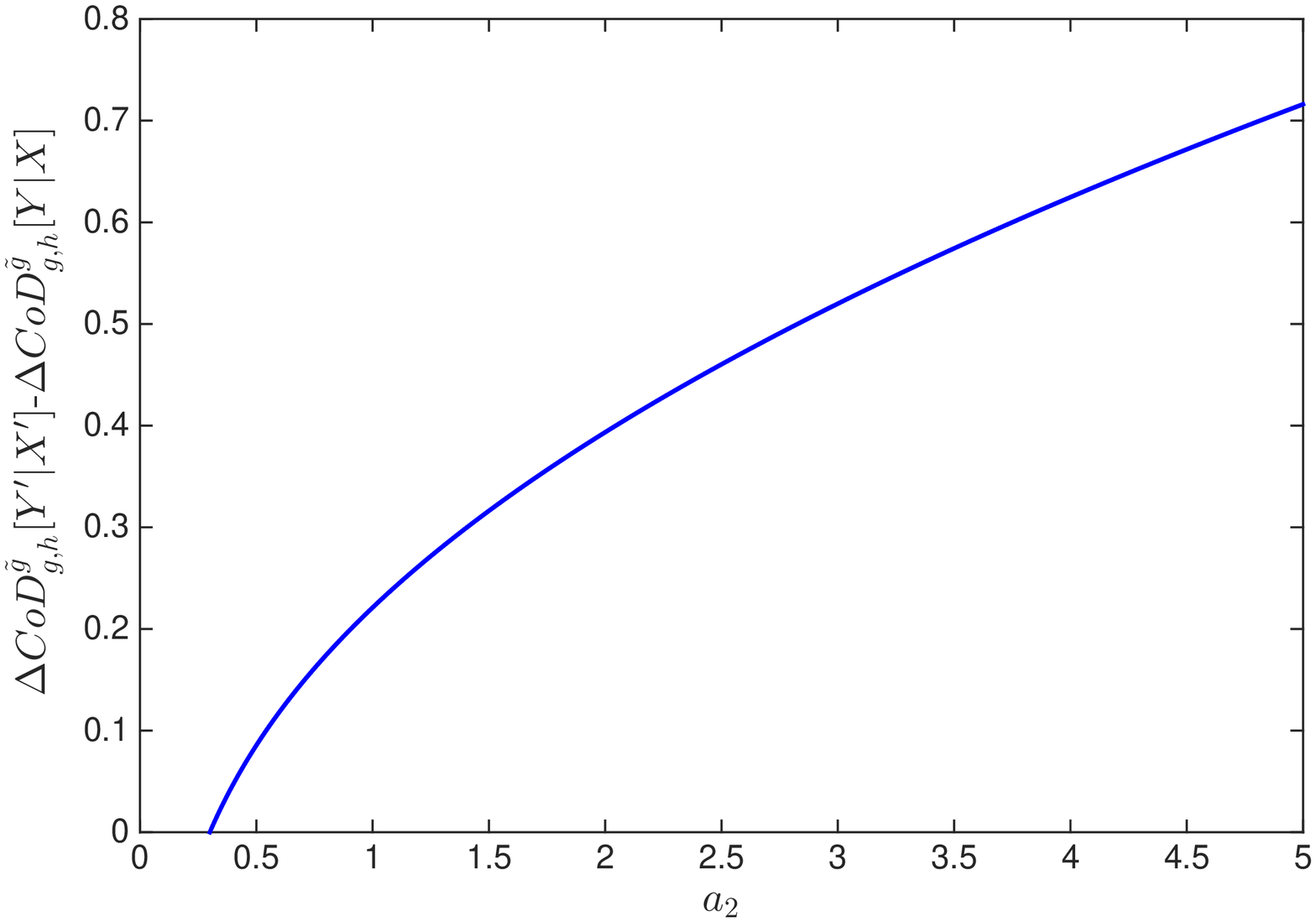}\label{figcontri4}}
	\caption{\footnotesize(a) Plot of $\Delta\mbox{{\rm CoD}}_{g,h}[Y|X]$ and $\Delta\mbox{{\rm CoD}}_{g,h}[Y'|X']$ on $\theta\geq1$. (b) Plot of $\Delta\mbox{{\rm CoD}}_{g,h}[Y|X]$ on $\gamma>0$. (c) Plot of $\Delta\mbox{{\rm CoD}}_{g,h}[Y|X]$ and $\Delta\mbox{{\rm CoD}}_{g,h'}[Y'|X']$ on $u_{g}\in[0.6,0.99]$. (d) Plot of $\Delta\mbox{{\rm CoD}}_{g,h}^{\tilde{g}}[Y'|X']-\Delta\mbox{{\rm CoD}}_{g,h}^{\tilde{g}}[Y|X]$ for different values of the shape parameter $a_2\geq a_1$.}\label{figcontri}
\end{figure}


The next example supports our comparison results for the distortion risk contribution measures.
\begin{Exa}[Distortion risk contribution measures]In this example,
we assume that the distortion functions applied to $Y$ and $Y'$ are of the form of a power function.
\begin{itemize}
\item [(a)] Suppose that $Y\sim\Gamma(a_1,b_1)$ and $Y'\sim\Gamma(a_2,b_2)$ with $(a_1,b_1)=(0.3,1)$ and $(a_2,b_2)=(2,1)$.
    Thus, it holds that $Y\leq_{\rm disp}Y'$. Let $u_{g}=0.8$ and $h(p)=p^{0.4}$, for $p\in[0,1]$.
    Figure \ref{figcontri1} displays the plots of $\Delta\mbox{{\rm CoD}}_{g,h}[Y|X]$ and $\Delta\mbox{{\rm CoD}}_{g,h}[Y'|X']$ on $\theta\geq1$, from which one can observe that $\Delta\mbox{{\rm CoD}}_{g,h}[Y|X]\leq \Delta\mbox{{\rm CoD}}_{g,h}[Y'|X']$ for any fixed $\theta$, and both of them are increasing with respect to `$\prec$'.
    This supports the result of Theorem \ref{thedispCON1}(i).
\item [(b)] Let $Y\sim\Gamma(0.2,1)$, $u_{g}=0.9$, and $\theta=2$.
    It is clear that $Y$ is DFR.
    The value of $\Delta\mbox{{\rm CoD}}_{g,h}[Y|X]$ is plotted in Figure \ref{figcontri2} for different distortion functions applied to $Y$.
    It is straightforward to observe that $\Delta\mbox{{\rm CoD}}_{g,h}[Y|X]$ is decreasing with respect to $\gamma$, which verifies Theorem \ref{theDFRCo}(i).
\item [(c)] Let $h(p)=p^{\gamma_1}$, $h'(p)=p^{\gamma_2}$, $C$ with parameter $\theta_1$ and $C'$ with parameter $\theta_2$. Set $\gamma_1=3$, $\gamma_2=2$, $\theta_1=2$, $\theta_2=3$, $Y\sim\Gamma(0.2,1)$, and $Y'\sim\Gamma(2,1)$.
    Figure \ref{figcontri3} plots $\Delta\mbox{{\rm CoD}}_{g,h}[Y|X]$ and $\Delta\mbox{{\rm CoD}}_{g,h'}[Y'|X']$ on $u_{g}\in[0.6,0.99]$.
    We observe that both $\Delta\mbox{{\rm CoD}}_{g,h}[Y|X]$ and $\Delta\mbox{{\rm CoD}}_{g,h'}[Y'|X']$ are increasing with respect to $u_{g}$,
    and  $\Delta\mbox{{\rm CoD}}_{g,h}[Y|X]\leq\Delta\mbox{{\rm CoD}}_{g,h'}[Y'|X']$ for any fixed $u_{g}$, which validates the result of Theorem \ref{thetotaldisp}.
\item [(d)] Assume that  $u_{g}=0.9$, $u_{\tilde{g}}=0.8$, $h(p)=p^{2}$, $\theta=2$, $Y\sim \Gamma(a_1,1)$, and $Y'\sim \Gamma(a_2,1)$ with $a_2>0$.
    The difference function between $\Delta\mbox{{\rm CoD}}_{g,h}^{\tilde{g}}[Y'|X']$ and $\Delta\mbox{{\rm CoD}}_{g,h}^{\tilde{g}}[Y|X]$ is plotted in Figure \ref{figcontri4}, which is always negative for all $a_2\geq a_1=0.3$.
    Thus, the result of Theorem \ref{medthedisp1} is validated.
\end{itemize}
\end{Exa}

\begin{figure}[htbp!]
	\centering
	\subfigure[]{\includegraphics[width=.48\textwidth, height=0.3\textheight]{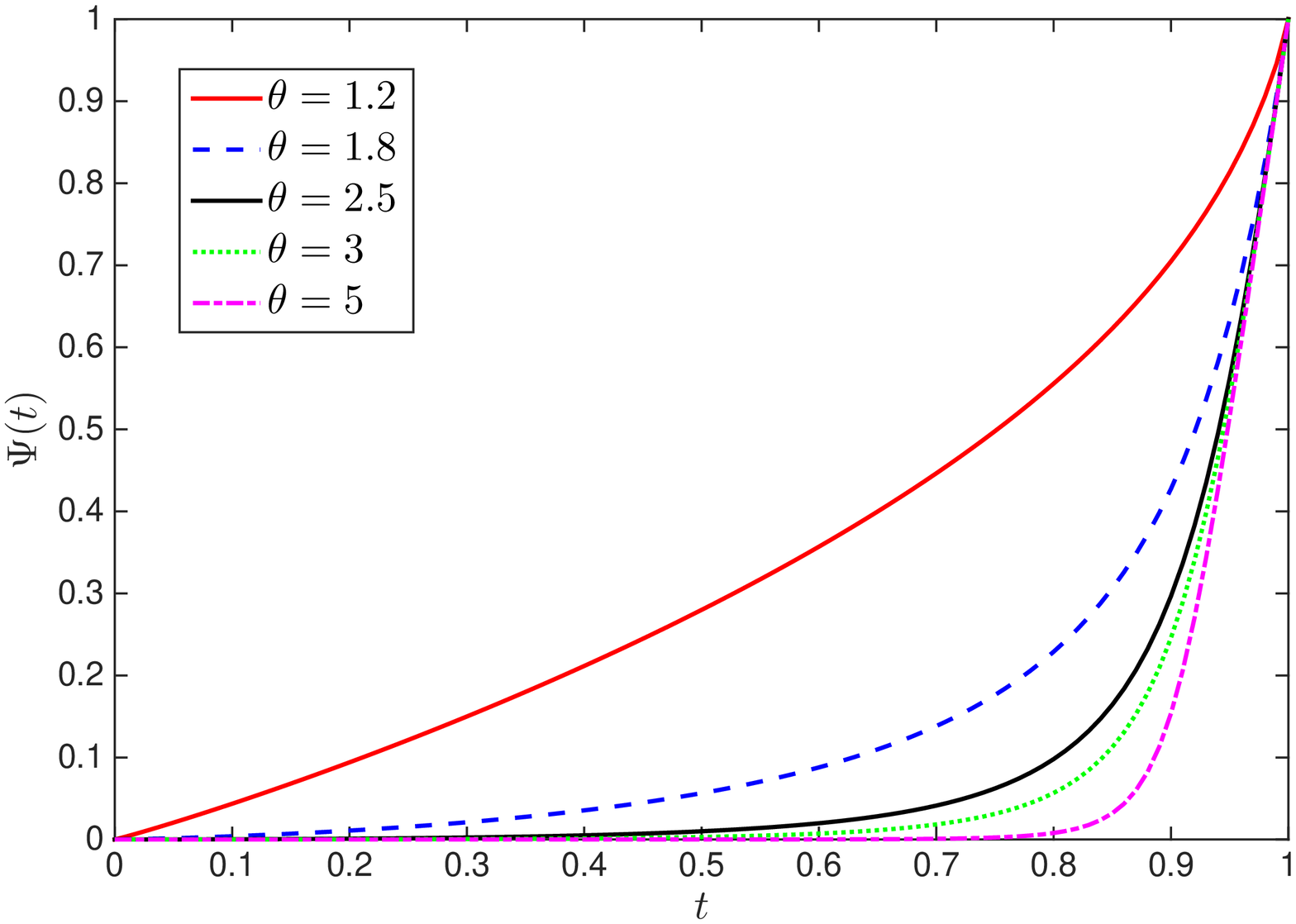}\label{figew1}}
	\subfigure[]{\includegraphics[width=.48\textwidth, height=0.3\textheight]{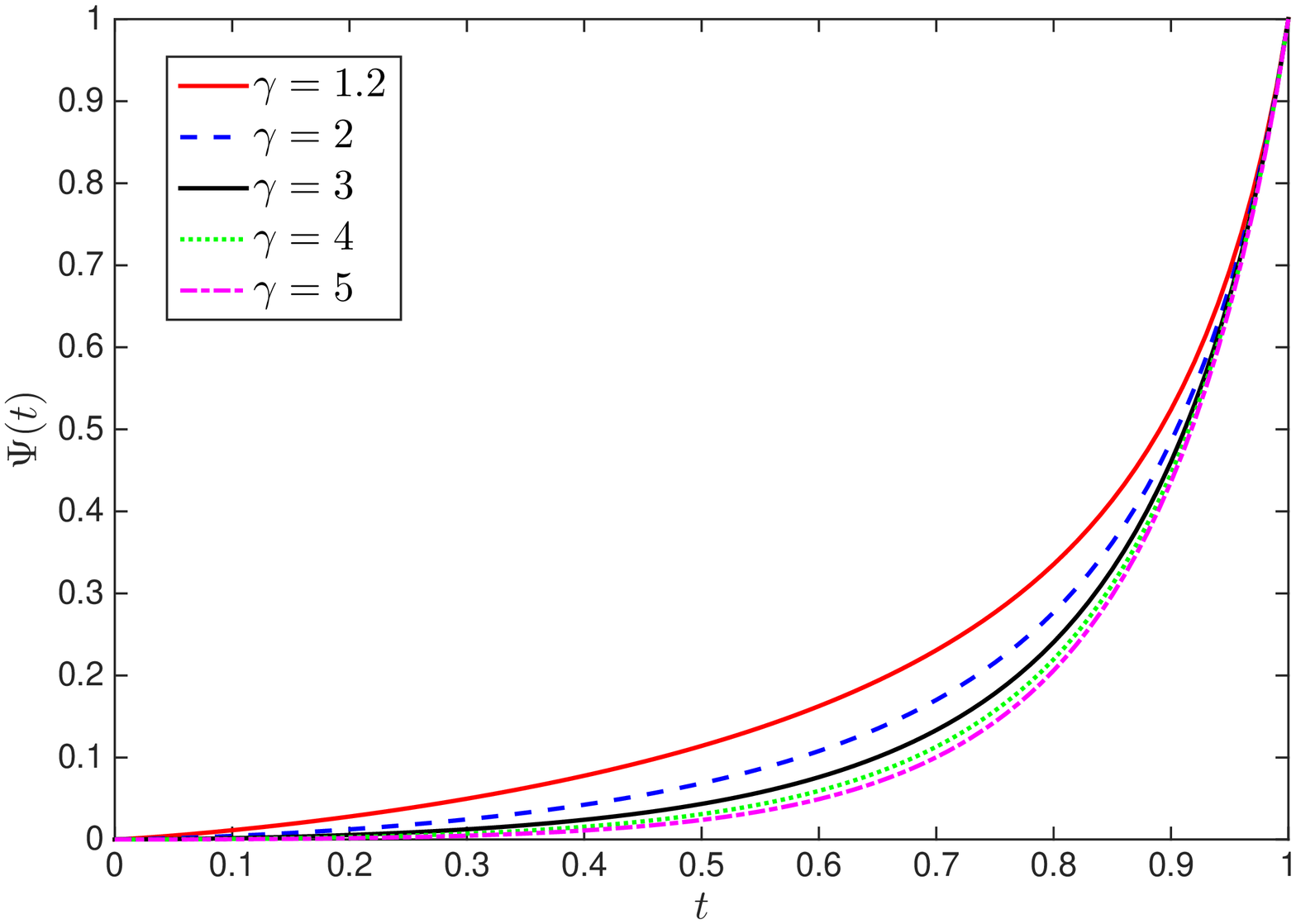}\label{figew2}}
	\caption{\footnotesize(a) Plot of $\Psi(t)$ on $t\in[0,1]$ for different values of $\theta$. (b) Plot of $\Psi(t)$ on $t\in[0,1]$ for different values of $\gamma$.}\label{figew}
\end{figure}

Next, we present an example to illustrate the condition in Theorem \ref{theewCON1}.
\begin{Exa} Assume that $h(p)=1-(1-p)^{\gamma}$ for $\gamma>1$.
Let $C$ be the Gumbel copula with dependence parameter $\theta>1$.
It is easy to verify that $h(p)$ is concave and $\widebar{h}(p)=p^{\gamma}$.
Observe that
\begin{equation*}
\Psi(t):=\widebar{h}(A(\widebar{h}^{-1}(t)))=\left[\frac{t^{\frac{1}{\gamma}}-C(u_{g},t^{\frac{1}{\gamma}})}{1-u_{g}}\right]^{\gamma}.
\end{equation*}
\begin{itemize}
\item [(a)] Set $u_{g}=0.9$ and $\gamma=1.1$.
    Figure \ref{figew1} plots $\Psi(t)$ on $t\in[0,1]$ under different values of $\theta=1.2,1.8,2.5,3,5$, which indicates the convexity of $\Psi(t)$.
\item [(b)] Set $u_{g}=0.9$ and $\theta=1.5$.
    Figure \ref{figew2} plots $\Psi(t)$ on $t\in[0,1]$ under different values of $\gamma=1.2,2,3,4,5$, from which one can observe the convexity of $\Psi(t)$.
\end{itemize}
\end{Exa}

\begin{figure}[htbp!]
	\centering{\includegraphics[width=.65\textwidth, height=0.3\textheight]{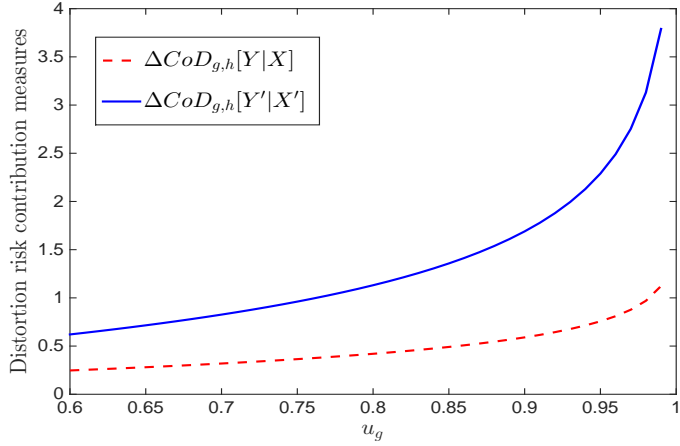}}
	\caption{\footnotesize Plot of $\Delta\mbox{{\rm CoD}}_{g,h}[Y|X]$ and $\Delta\mbox{{\rm CoD}}_{g,h}[Y'|X']$ for $u_{g}\in[0.6,0.99]$.}\label{figewcon}
\end{figure}

The following example illustrates Theorem \ref{theewCON1}.
\begin{Exa} Assume that $h(p)=1-(1-p)^{2}$, $\theta=1.5$, $Y\sim W(1,2)$, and $Y\sim W(1,1)$.
Clearly, it holds that $Y\leq_{\rm ew}Y'$ but $Y\nleq_{\rm disp}Y'$ nor $Y\ngeq_{\rm disp}Y'$ \citep[see Example 24 in][]{sordo2018IME}.
As displayed in Figure \ref{figewcon}, $\Delta\mbox{{\rm CoD}}_{g,h}[Y|X]\leq \Delta\mbox{{\rm CoD}}_{g,h}[Y'|X']$ for $u_{g}\in[0.6,0.99]$, which shows the effectiveness of Theorem \ref{theewCON1}.
\end{Exa}

Next, we present a numerical example to show the effectiveness of Theorem \ref{thepair1}.
\begin{Exa} Let $C$ be the Gumbel copula with dependence parameter $\theta=2$.
Assume that $g(t)=t^{0.3}$, $X\sim\Gamma(0.5,1)$, $Y\sim\Gamma(1.5,1)$, and $h(p)=p^{\gamma_2}$ for $\gamma_2>0$.
It is easy to verify that $X\leq_{\rm st}Y$ and $X\leq_{\rm disp}Y$.
Moreover, one can calculate that $u_{g}^{X}=0.9714>u_{g}^{Y}=0.9599$.
Figure \ref{figpair1} displays $\mbox{{\rm CoD}}_{g,h}[Y|X]$ and $\mbox{{\rm CoD}}_{g,h}[X|Y]$ for $\gamma_2>0$, and Figure \ref{figpair2} plots $\Delta\mbox{{\rm CoD}}_{g,h}[Y|X]$ and $\Delta\mbox{{\rm CoD}}_{g,h}[X|Y]$ for $\gamma_2>0$.
Obviously, $\mbox{{\rm CoD}}_{g,h}[Y|X]\geq\mbox{{\rm CoD}}_{g,h}[X|Y]$ and $\Delta\mbox{{\rm CoD}}_{g,h}[Y|X]\geq\Delta\mbox{{\rm CoD}}_{g,h}[X|Y]$ for $\gamma_2>0$.
Therefore, the results of Theorem \ref{thepair1}(i) and Theorem \ref{thepair1}(v) are supported.
\begin{figure}[htbp!]
	\centering
	\subfigure[]{\includegraphics[width=.48\textwidth, height=0.3\textheight]{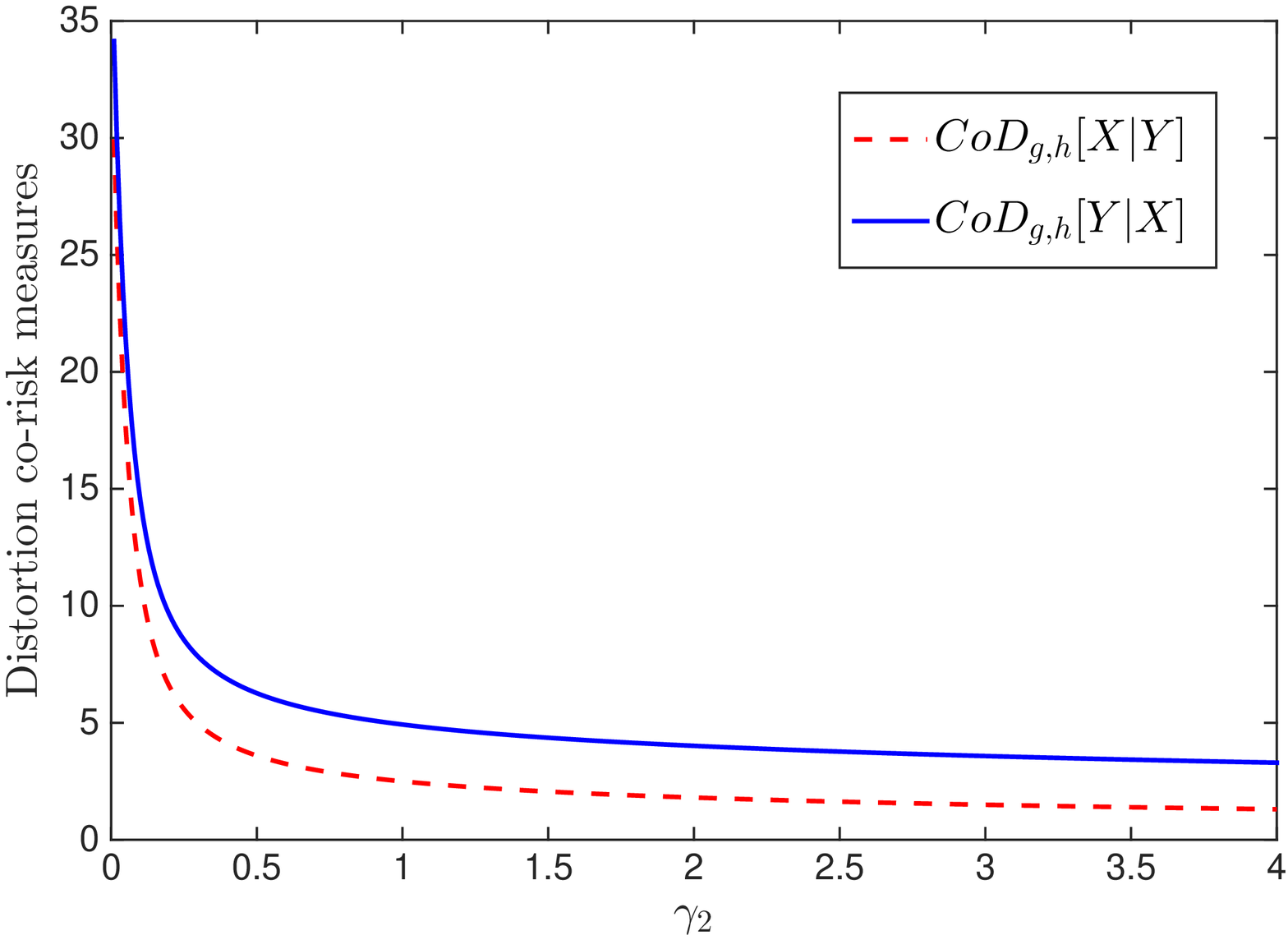}\label{figpair1}}
	\subfigure[]{\includegraphics[width=.48\textwidth, height=0.3\textheight]{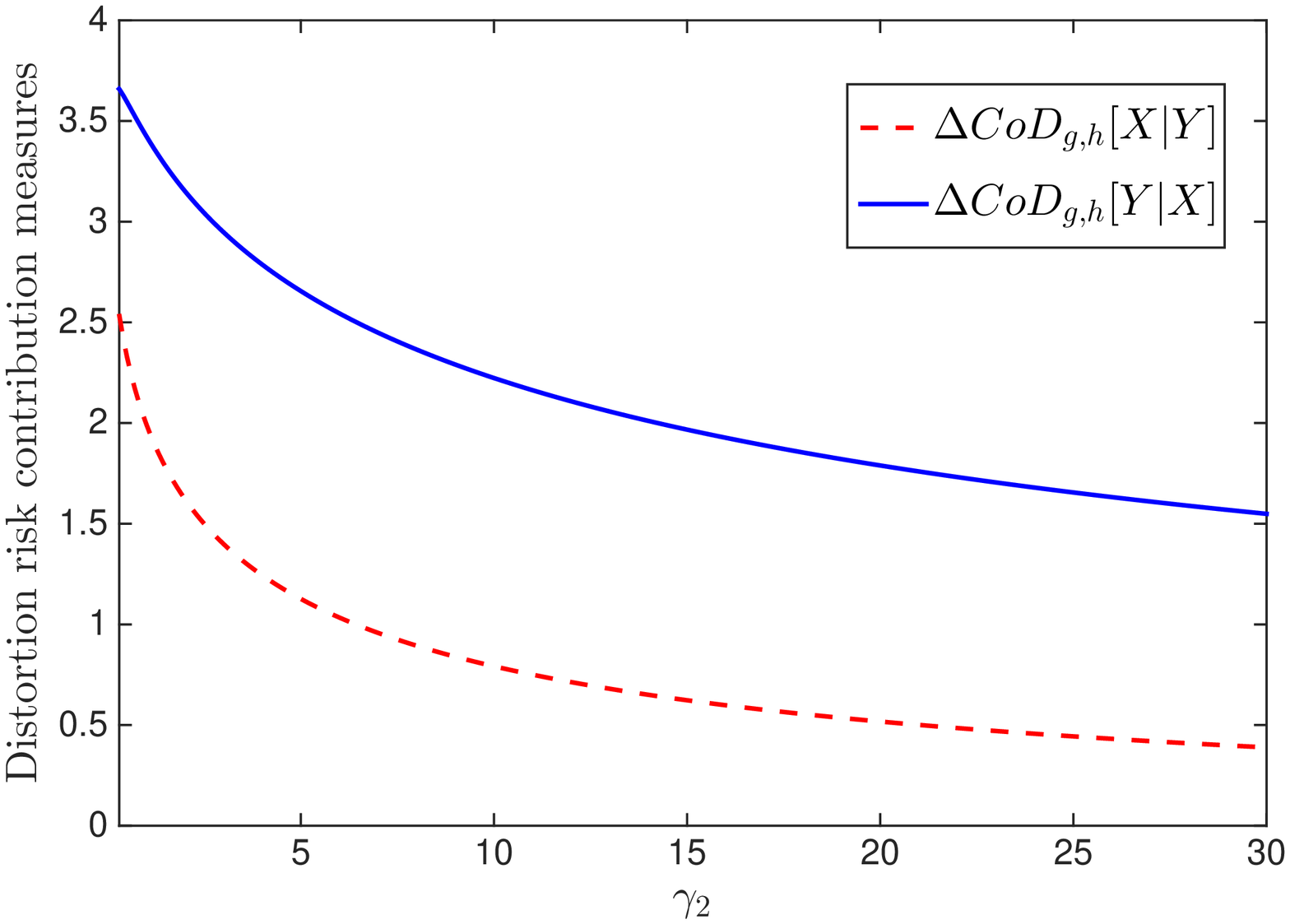}\label{figpair2}}
	\caption{\footnotesize(a) Plot of $\mbox{{\rm CoD}}_{g,h}[Y|X]$ and $\mbox{{\rm CoD}}_{g,h}[X|Y]$ for $\gamma_2>0$. (b) Plot of $\Delta\mbox{{\rm CoD}}_{g,h}[Y|X]$ and $\Delta\mbox{{\rm CoD}}_{g,h}[X|Y]$ for $\gamma_2>0$.}\label{figpair}
\end{figure}
\end{Exa}

\subsection{The Farlie-Gumbel-Morgenstern Copula}
The Farlie-Gumbel-Morgenstern (FGM) copula is defined as
\begin{equation*}
C_{\alpha}(u,v)=uv\left[1+\alpha(1-u)(1-v)\right], \quad\mbox{$-1\leq\alpha\leq1$}.
\end{equation*}
If $\theta=0$, then $C_{\theta}$ reduces to the independence copula.
Furthermore, $C_{\alpha}(u,v)$ is RR$_2$ [TP$_2$] for $\alpha\in[-1,0)$ [$\alpha\in[0,1]$] and $\alpha_{1}\leq\alpha_2$ implies that $C_{\alpha_1}\prec C_{\alpha_2}$.
For more details on its properties, we refer to \cite{Joe1997} and \cite{Nelsen2007}.

 \begin{figure}[htbp!]
	\centering
	\subfigure[]{\includegraphics[width=.48\textwidth, height=0.3\textheight]{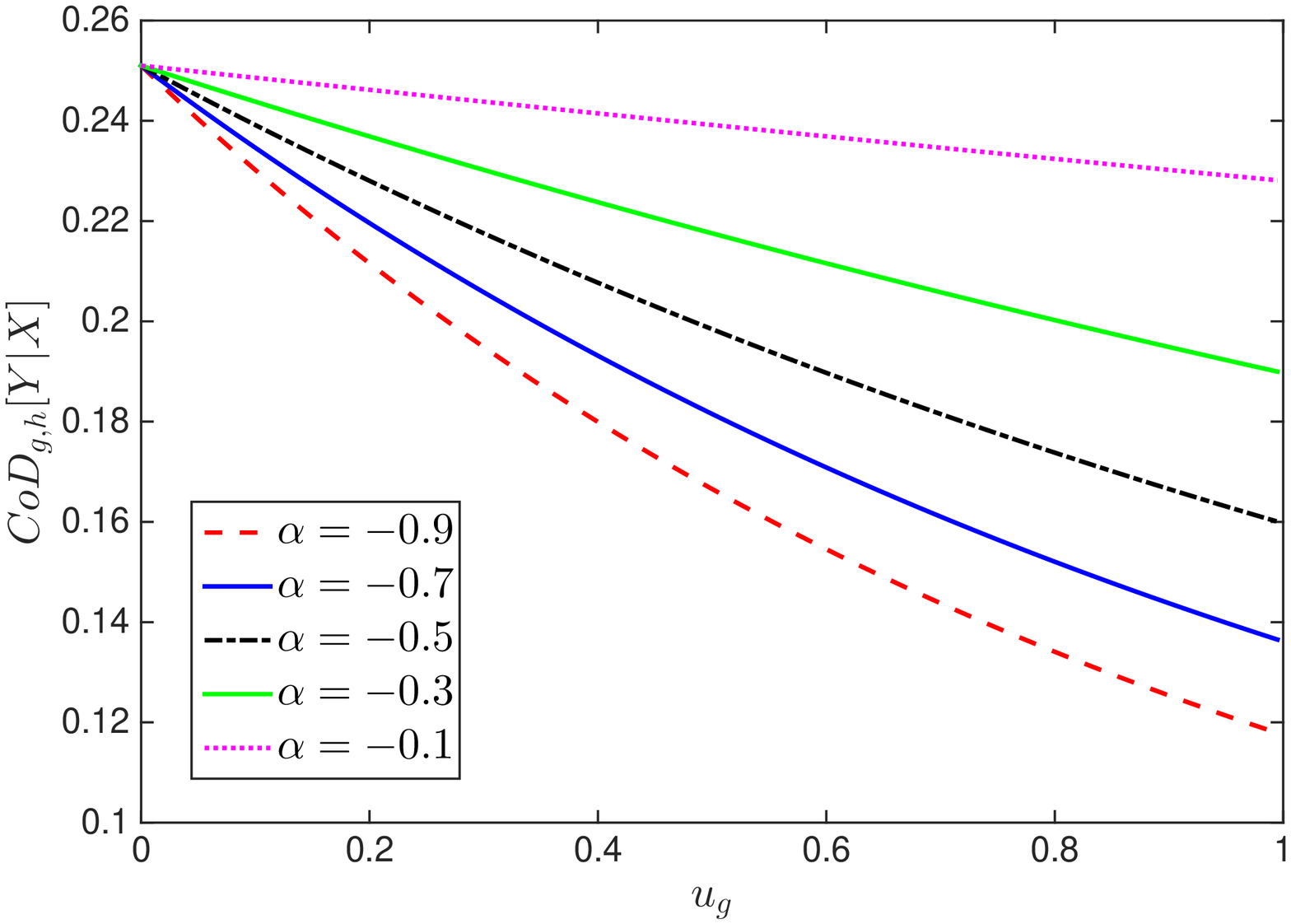}\label{negfig1}}
	\subfigure[]{\includegraphics[width=.48\textwidth, height= 0.3\textheight]{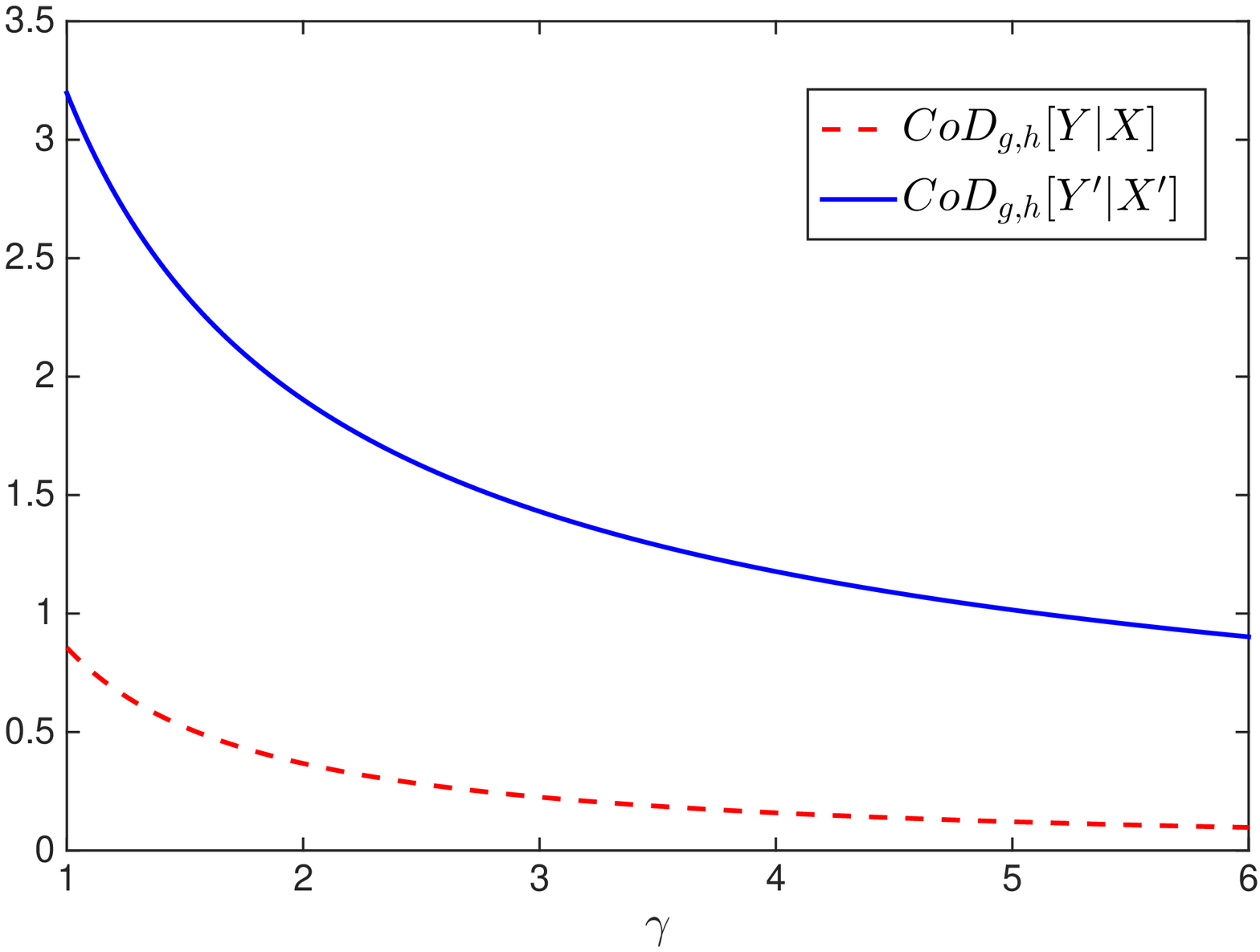}\label{negfig2}}
	\subfigure[]{\includegraphics[width=.48\textwidth, height= 0.3\textheight]{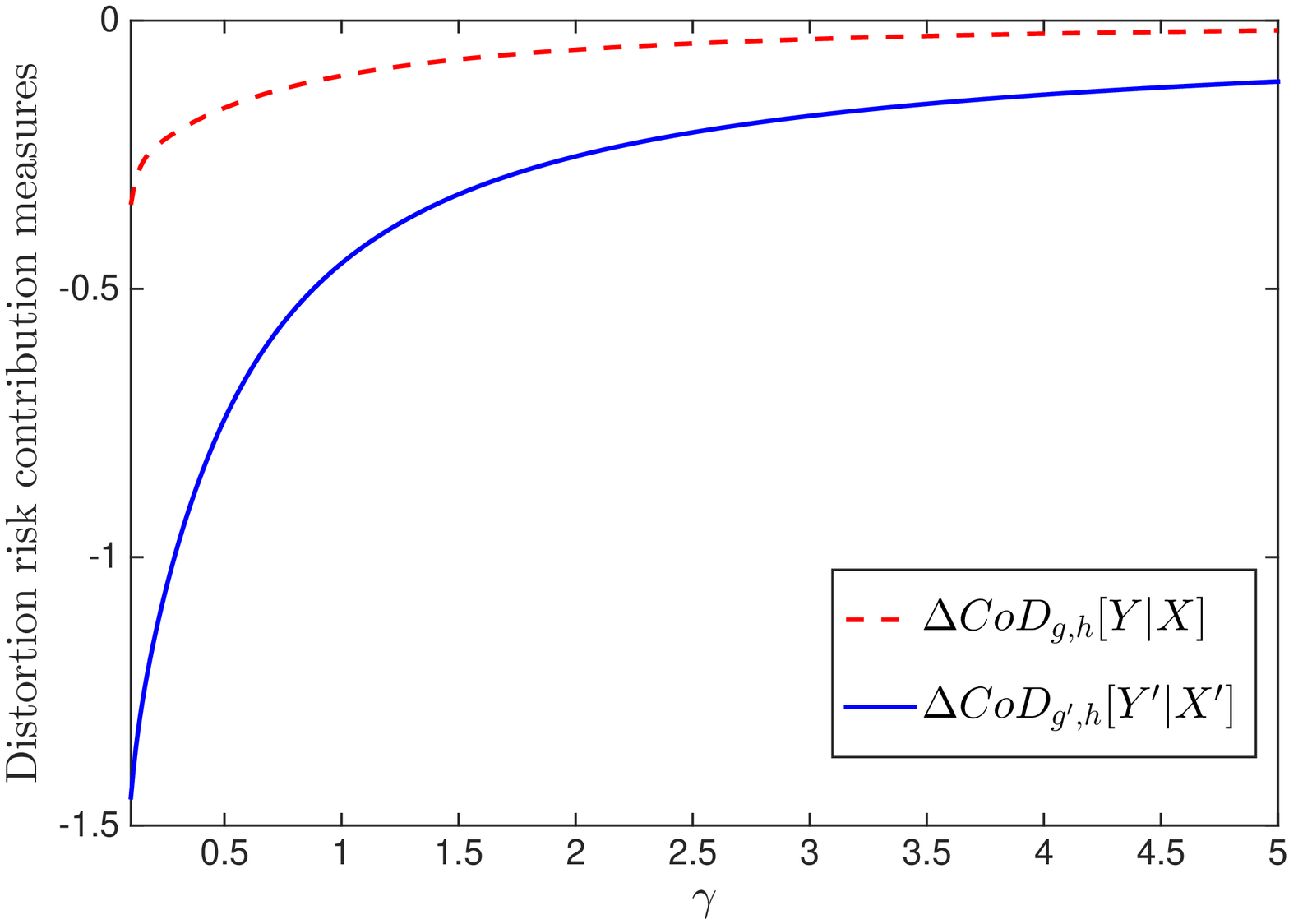}\label{negfig3}}
	\subfigure[]{\includegraphics[width=.48\textwidth, height= 0.3\textheight]{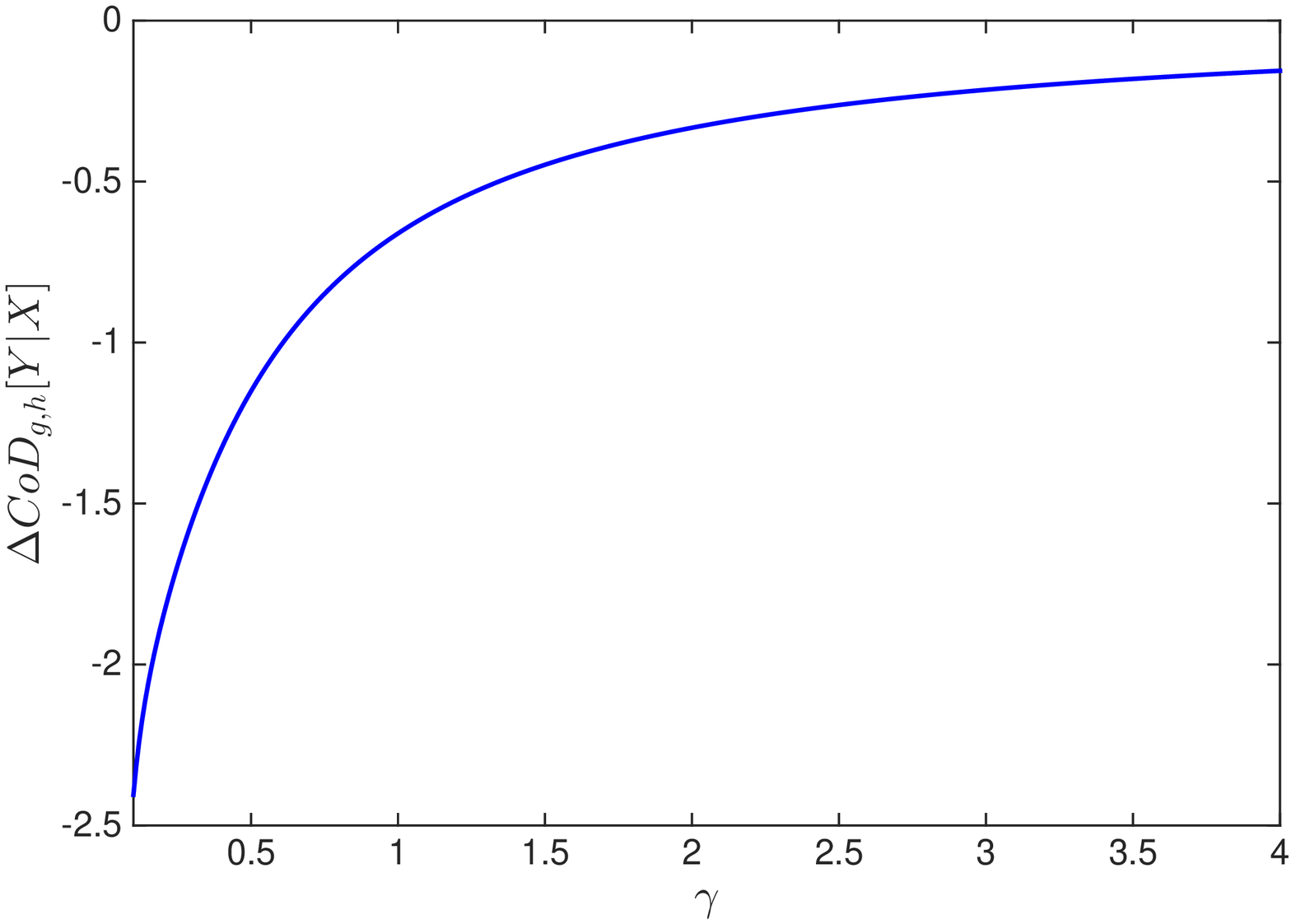}\label{negfig4}}
	\caption{\footnotesize(a) Plot of $\mbox{{\rm CoD}}_{g,h}[Y|X]$ for $u_{g}\in[0,1)$. (b) Plot of $\Delta\mbox{{\rm CoD}}_{g,h}[Y|X]$ and $\Delta\mbox{{\rm CoD}}_{g,h}[Y'|X']$ for $\gamma\geq1$. (c) Plot of $\Delta\mbox{{\rm CoD}}_{g,h}[Y|X]$ and $\Delta\mbox{{\rm CoD}}_{g',h}[Y'|X']$ for $\gamma>0$. (d) Plot of $\Delta\mbox{{\rm CoD}}_{g,h}[Y|X]$ for $\gamma>0$.}\label{negfig} 
\end{figure}

The following examples show the effectiveness of Theorem \ref{theg11}(ii), Theorem \ref{theicxco1}(ii), Theorem \ref{thedispCON8}(ii), Theorem \ref{theDFRCo}(ii), and Theorem \ref{thepair1} under the negative dependence characterized by the FGM copula.

\begin{figure}[htbp!]
	\centering
	\subfigure[]{\includegraphics[width=.48\textwidth, height=0.3\textheight]{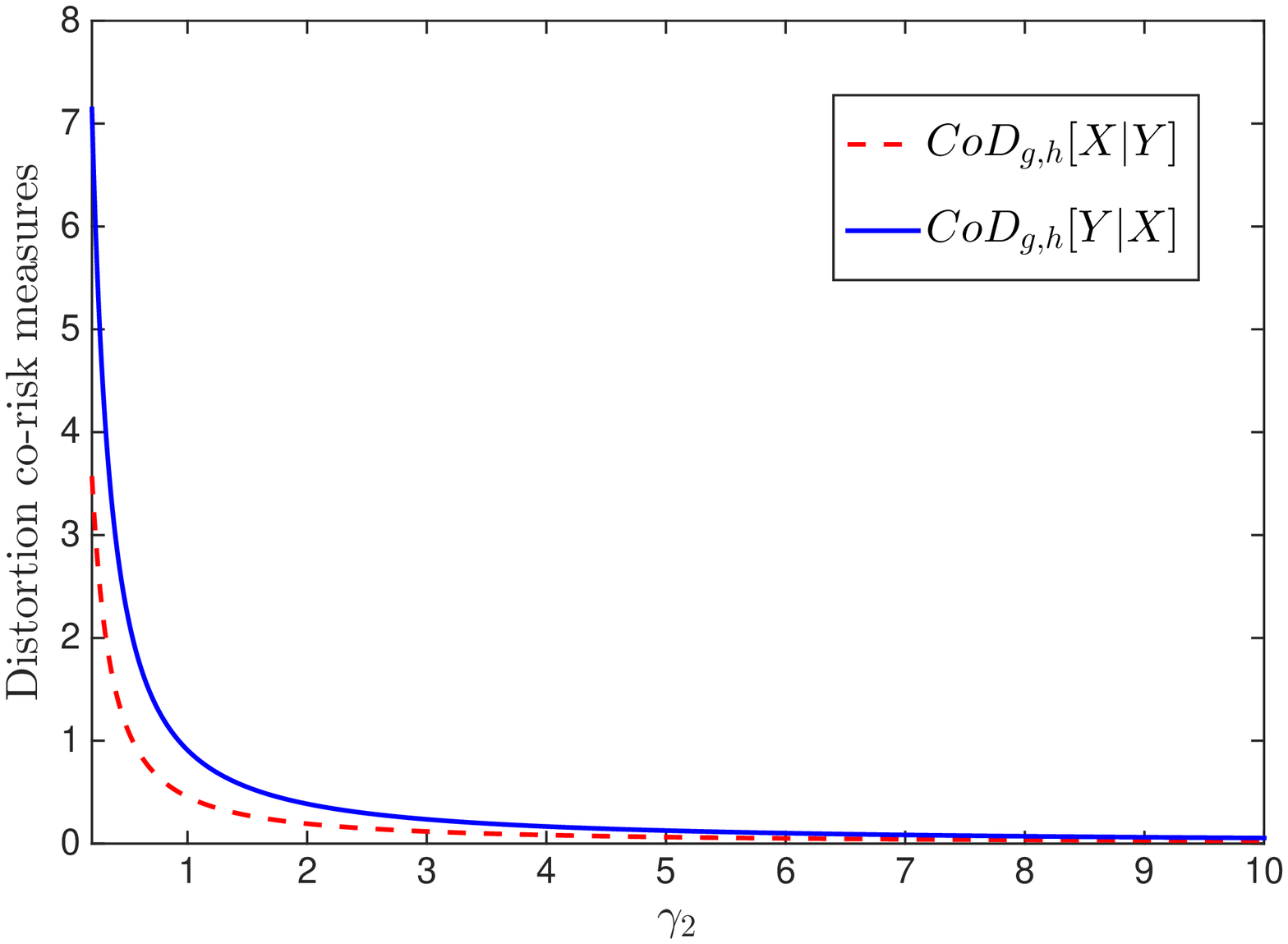}\label{fignegpair1}}
	\subfigure[]{\includegraphics[width=.48\textwidth, height=0.3\textheight]{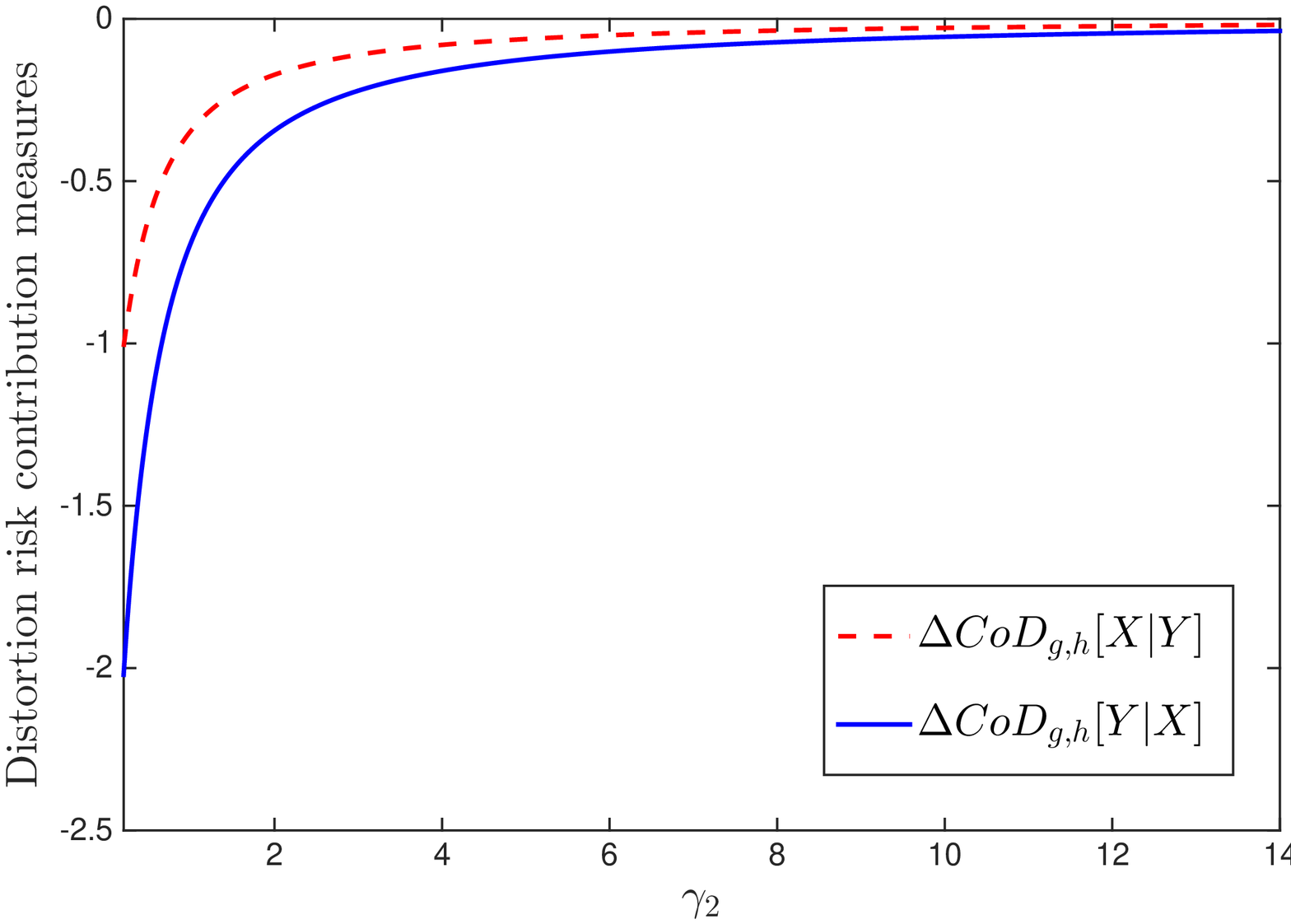}\label{fignegpair2}}
	\caption{\footnotesize(a) Plot of $\mbox{{\rm CoD}}_{g,h}[Y|X]$ and $\mbox{{\rm CoD}}_{g,h}[X|Y]$ for $\gamma_2>0$. (b) Plot of $\Delta\mbox{{\rm CoD}}_{g,h}[Y|X]$ and $\Delta\mbox{{\rm CoD}}_{g,h}[X|Y]$ for $\gamma_2>0$.}\label{fignegpair}
\end{figure}

\begin{Exa}
\begin{itemize}
\item [(a)] Set $Y\sim\Gamma(0.8,2)$ and $h(p)=p^{5}$ for $p\in[0,1]$.
    Figure \ref{negfig1} displays $\mbox{{\rm CoD}}_{g,h}[Y|X]$ on $u_{g}\in[0,1)$ for different values of the dependence parameter $\alpha=-0.9,-0.7,-0.5,-0.3,-0.1$.
    One readily observes that $\mbox{{\rm CoD}}_{g,h}[Y|X]$ is decreasing with respect to $u_{g}$ for any fixed $\alpha$,
    while it is increasing in $\alpha$ for any fixed $u_{g}$.
    This agrees with the result of Theorem \ref{theg11}(ii).
\item [(b)] Set $Y\sim\Gamma(0.8,2)$, $Y'\sim\Gamma(1.8,2)$, $u_{g}=0.95$, $\alpha_1=-0.9$, and $\alpha_2=-0.3$.
    Let $h(p)=p^{\gamma}$ for $\gamma\geq1$ and $p\in[0,1]$, which means that $h$ is increasing and convex.
    The values of $\Delta\mbox{{\rm CoD}}_{g,h}[Y|X]$ and $\Delta\mbox{{\rm CoD}}_{g,h}[Y'|X']$ are plotted in Figure \ref{negfig2} for $\gamma\geq1$, from which it is clear that $\Delta\mbox{{\rm CoD}}_{g,h}[Y|X]\leq\Delta\mbox{{\rm CoD}}_{g,h}[Y'|X']$ for $\gamma\geq1$.
    Thus, the result of Theorem \ref{theicxco1}(ii) is validated.
\item [(c)] Suppose that $Y\sim\Gamma(0.6,1)$, $Y'\sim\Gamma(1.2,1)$, $u_{g}=0.95$, $u_{g'}=0.9$, $\alpha=-0.3$, and $\alpha'=-0.9$.
    It is plain that $Y\leq_{\rm disp}Y'$, $u_{g}\geq u_{g'}$, and $C'\prec C$.
    As observed from Figure \ref{negfig3}, $\Delta\mbox{{\rm CoD}}_{g,h}[Y|X]\geq\Delta\mbox{{\rm CoD}}_{g',h}[Y'|X']$ for $\gamma>0$,
    which illustrates Theorem \ref{thedispCON8}(ii).
    \item [(d)] Set $\alpha=-0.8$, $Y\sim\Gamma(0.8,2)$, and $u_{g}=0.95$.
    Thus, $Y$ is DFR.
    Let $h(p)=p^{\gamma}$ for $\gamma>0$.
    As shown in Figure \ref{negfig4}, the value of $\Delta\mbox{{\rm CoD}}_{g,h}[Y|X]$ is increasing with respect to $\gamma>0$, which validates the theoretical finding of Theorem \ref{theDFRCo}(ii).
\end{itemize}
\end{Exa}

Finally, a numerical example is provided to show the effectiveness of Theorem \ref{thepair1} under the case of negative dependence.
\begin{Exa}
Let $C$ be the FGM copula with dependence parameter $\alpha=-0.8$.
Assume that $g(p)=p^{0.2}$, $X\sim\Gamma(0.8,1)$, $Y\sim\Gamma(0.8,2)$, and $h(p)=p^{\gamma_2}$ for $\gamma_2>0$.
It is easy to verify that $X\leq_{\rm hr}Y$ and thus $X\leq_{\rm st~[disp]}Y$ since both $X$ and $Y$ are DFR.
Moreover, one can calculate that $u_{g}^{X}=u_{g}^{Y}=0.9937$.
Figure \ref{fignegpair1} displays $\mbox{{\rm CoD}}_{g,h}[Y|X]$ and $\mbox{{\rm CoD}}_{g,h}[X|Y]$ for $\gamma_2>0$, and Figure \ref{fignegpair2} plots $\Delta\mbox{{\rm CoD}}_{g,h}[Y|X]$ and $\Delta\mbox{{\rm CoD}}_{g,h}[X|Y]$ for $\gamma_2>0$.
Note that $\mbox{{\rm CoD}}_{g,h}[Y|X]\geq\mbox{{\rm CoD}}_{g,h}[X|Y]$ while $\Delta\mbox{{\rm CoD}}_{g,h}[Y|X]\leq\Delta\mbox{{\rm CoD}}_{g,h}[X|Y]$ for all $\gamma_2>0$, and thus the results of Theorems \ref{thepair1}(ii) and \ref{thepair1}(iv) are validated.
\end{Exa}

\section{Conclusions}\label{sec:conclusions}
We have introduced the rich classes of conditional distortion (CoD) risk measures and distortion risk contribution ($\Delta$CoD) measures,
which include many of the existing measures proposed in the academic literature related to systemic risk as special cases.
We have analyzed their properties and representations. 
We have given sufficient conditions for two random vectors to be ordered by the proposed measures,
which are expressed using the conventional stochastic order, the increasing convex [concave] order, the dispersive order, 
and the excess wealth order of marginals,
under explicit assumptions of positive or negative dependence, distortion functions, and threshold quantiles.
Numerical examples have been provided to illustrate the validity of our theoretical findings.
This work is the second in a triplet of papers on systemic risk by the same authors.
In \cite{DLZorder2018a}, we introduce and investigate some new stochastic orders that can be applied in the context of systemic risk evaluation,
while the present article introduces conditional distortion risk measures applicable in this context.
In a third (forthcoming) paper, we will combine the results of both papers to attribute systemic risk to the different participants
in a given risky system.

\section*{Acknowledgements}
Jan Dhaene acknowledges the financial support of the Research Foundation Flanders (FWO) under
grant GOC3817N.
Roger Laeven acknowledges the financial support of the Netherlands
Organization for Scientific Research under grant NWO VIDI.
Yiying Zhang acknowledges
the financial support and nice working place from the Actuarial Research Group at KU Leuven
and the Amsterdam Center of Excellence in Risk and Macro Finance at the University of Amsterdam during his visit.

\bibliography{CoRisk}

\end{document}